\documentclass[11pt]{article}

\usepackage{amsmath, amsfonts, amssymb,latexsym,wasysym,graphicx}
\input epsf.sty

\newtheorem{Theorem}{Theorem}[section]
\newtheorem{Lemma}{Lemma}[section]

\newtheorem{Definition}[Lemma]{Definition}

\newcommand{\BEQ}{\begin{equation}}     
\newcommand{\BEA}{\begin{eqnarray}}
\newcommand{\BD}{\begin{displaymath}}
\newcommand{\EEQ}{\end{equation}}       
\newcommand{\EEA}{\end{eqnarray}}
\newcommand{\ED}{\end{displaymath}}

\newcommand{\del}{\delta}
\newcommand{\Del}{\Delta}
\newcommand{\eps}{\varepsilon}          

\newcommand{\Tr}{{\mathrm{Tr}}}

\newcommand{\R}{\mathbb{R}}
\newcommand{\C}{\mathbb{C}}
\newcommand{\Z}{\mathbb{Z}}

\newcommand{\II}{{\rm i}}               
\renewcommand{\Re}{{\rm Re\ }}          
\renewcommand{\Im}{{\rm Im\ }}          
\newcommand{\half}{{1\over 2}}          
\renewcommand{\vec}[1]{\boldsymbol{#1}} 

\catcode`\@=11
\def\numberbysection{\@addtoreset{equation}{section}
        \def\theequation{\thesection.\arabic{equation}}}
\numberbysection

\begin{document}

\vspace*{1.5cm}
\begin{center}
{\Large \bf Dynamical invariance for random matrices}

\end{center}

\vspace{2mm}
\begin{center}
{\bf  J\'er\'emie Unterberger}
\end{center}

\vspace{2mm}
\begin{quote}

\renewcommand{\baselinestretch}{1.0}
\footnotesize
{We consider a general Langevin dynamics for the one-dimensional N-particle Coulomb
gas with confining potential $V$ at temperature $\beta$. These dynamics describe for $\beta=2$ the time evolution of the
eigenvalues of 
$N\times N$ random Hermitian matrices.  The equilibrium partition function -- equal to the
normalization constant of the Laughlin wave function in fractional quantum Hall
effect -- is known to satisfy an infinite number of constraints 
called Virasoro or loop constraints. We introduce here a dynamical generating function
on the space of random trajectories which satisfies a large class of constraints of
geometric origin. We focus in this article on a subclass induced by the invariance
under the Schr\"odinger-Virasoro algebra.
}
\end{quote}

\vspace{4mm}


 \medskip
 

\tableofcontents



\section{Introduction}


Let us start with a short preliminary discussion of the model (\S 0.1) and
of the well-known equilibrium Virasoro constraints (\S 0.2). For a presentation
of our results, the reader familiar with these may skip directly to \S 0.3.

\subsection{Dyson's Brownian motion}


We consider the following Langevin dynamics  \cite{Dys,For,Tao} for $N$ particles confined to a line
with positions $\{\lambda_i\}$, $i=1,\ldots,N$
\BEQ d\lambda_i=dB_i-\frac{\partial W}{\partial\lambda_i} dt=dB_i+
\left( \sum_{j\not=i} \frac{\beta}{\lambda_i-\lambda_j}-V'(\lambda_i) \right)dt  \label{eq:L} \EEQ
where:

\noindent (i) the noises $(B_1,\ldots,B_N)=(B_1(t),\ldots,B_N(t))$ are $N$ independent
Brownian motions; 

\noindent (ii) $W({\lambda_i})=-\frac{\beta}{2} \sum_{i,j\not=i} \log|\lambda_i-\lambda_j| +
\sum_i V(\lambda_i)$ is the sum of the electrostatic energy of a system of $N$ identically charged particles and of a one-body confining potential $V$.

In our convention, $d\langle B_i,B_i\rangle_t=2dt$. Then the probability distribution function ${\cal P}({\lambda_i};t)$ for the positions of the particles satisfies the Fokker-Planck equation,
\BEA \partial_t {\cal P} & =& \Del {\cal P}+\sum_i \frac{\partial}{\partial\lambda_i} \left(
\frac{\partial W}{\partial\lambda_i} {\cal P} \right)  \nonumber\\
&=& \sum_i \frac{\partial^2 {\cal P}}{\partial \lambda_i^2} -\beta\sum_{i,j\not=i} 
\frac{\partial}{\partial\lambda_i} \left( \frac{ {\cal P}}{\lambda_i-\lambda_j} \right) +
\sum_i \frac{\partial}{\partial\lambda_i} \left( V'(\lambda_i) {\cal P} \right) 
\label{eq:FP} \EEA
(with the usual normalization of Brownian motion one would get $\half \Del$ in the
above expression).

For $V$ growing sufficiently fast at $\infty$,  the
unique stationary measure is the Gibbs measure 
\BEQ {\cal P}_{eq}(\{\lambda_i\})=\frac{1}{Z_N(V)} \frac{e^{-W({\lambda_i})}}{N!} \, d\lambda=
\frac{1}{Z_N(V)} \frac{1}{N!} \prod_{i=1}^N e^{-V(\lambda_i)} \prod_{i,j>i} (\lambda_j-\lambda_i)^{\beta}\, d\lambda \label{eq:Peq} \EEQ
which may also be interpreted as a normalization constant for the celebrated
fractional quantum Hall effect Laughlin wave-function \cite{Lau}.

For $\beta=2$, the normalization constant is the partition function of the Hermitian ensemble with potential $V$,
$Z_N(V)=\int d{\bf M}\, e^{-\Tr V({\bf M})}$, for a suitable normalization of the measure $d{\bf M}$
on the space of $N\times N$ Hermitian matrices. In fact, the measure $\frac{1}{Z_N(V)}
e^{-\Tr V({\bf M})} \, d{\bf M}$ projects down by conjugation invariance to a measure on the spectrum $\{\lambda_i\}$ of ${\bf M}$ which is none other than ${\cal P}_{eq}(\{\lambda_i\})$.
Following Dyson \cite{Dys} who originally introduced this model, we may consider our dynamics to be the projection to the spectrum of a conjugation invariant random walk on the space of Hermitian matrices, $d{\bf M}=d{\bf B}-V'({\bf M}) dt$, where by assumption the linearly independent entries $\Big(\{{\bf B}_{ii}(t)\}_i; \{\Re {\bf B}_{ij}(t) \}_{i<j};$ 
$\{\Im {\bf B}_{ij}(t)\}_{i<j} \Big)$ are independent Brownian motions; in other words, 
$d{\bf B}(t)$, $t\ge 0$ are independent infinitesimal increments drawn from GUE distribution.

In order to highlight the connection with the equilibrium case, we also use two
equivalent  reformulations of (\ref{eq:L}). First
we have a  measure $\cal Q$ on the space of trajectories $(\{\lambda_i(t)\})_{t\ge 0}$,
absolutely continuous with respect to the Wiener measure ${\cal D}{\cal W}$,
formally, 
\BEA  && {\cal Q}(\{\lambda_i\})={\cal Q}(\{\lambda_i(t)\})_{t\ge 0})  = {\cal D}\lambda\,
\exp  \left( - \int dt\, \sum_i (\dot{\lambda}_i+\frac{\partial W}{\partial\lambda_i})^2
\right)
\nonumber\\  && \qquad \qquad =
{\cal D}\lambda\,
\exp \left( - \int dt\, \sum_i (\dot{\lambda}_i-\sum_{j\not=i} \frac{\beta}{\lambda_i-\lambda_j}+
V'(\lambda_i))^2 \right). \label{eq:Q} \EEA
This can be made rigorous by using a Girsanov transformation \cite{RevYor}, namely,
\BEQ {\cal Q}(\{\lambda_i\})={\cal D}{\cal W}(\lambda) \, \exp\left( -2\sum_i
\int \frac{\partial W}{\partial\lambda_i}(t)\,  d\lambda_i(t) \, - \sum_i \int dt\, 
(\frac{\partial W}{\partial\lambda_i})^2(t) \right), \label{eq:Girsanov} \EEQ
where the integral $\int \frac{\partial W}{\partial\lambda_i}(t)\,  d\lambda_i(t)$ is
an It\^o integral.

\medskip

Then, by an elementary Hubbard-Stratonovich transformation, we obtain a (mathematically
ill-defined) complex Gibbs
measure on the space of trajectories  $(\{\lambda_i(t)\})_{t\ge 0}$,
$(\{\mu_i(t)\})_{t\ge 0}$ of  the particles and of associated virtual particles with positions
$\{ \mu_i\}$, also confined on the line,

\BEA  && \overline{\cal Q}(\{\lambda_i\},\{\mu_i\})=\overline{\cal Q}(\{\lambda_i(t)\})_{t\ge 0}, \{\mu_i(t)\}_{t\ge 0})  \nonumber\\
&& = {\cal D}\lambda\, {\cal D}\mu\, 
\exp \left(-\sqrt{-1} \int dt\, \sum_i \mu_i (\dot{\lambda}_i+\frac{\partial W}{\partial\lambda_i}) -\int dt\, \sum_i \mu_i^2 \right) 
\nonumber\\  && \qquad  =
{\cal D}\lambda\, {\cal D}\mu\, 
\exp \left( -\sqrt{-1} \int dt\, \sum_i \mu_i (\dot{\lambda}_i-\sum_{j\not=i} \frac{\beta}{\lambda_i-\lambda_j}+
V'(\lambda_i)) - \int dt\, \sum_i \mu_i^2  \right). \nonumber\\
 \label{eq:Qbar} \EEA
 
The above formula is a reformulation of the initial coupled stochastic differential
equations in the Martin-Siggia-Rose formalism \cite{MSR}.
 The quantities ${\cal L}, \overline{\cal L}$ in the exponentials, 
\BEQ {\cal Q}(\{\lambda_i\}) \equiv {\cal D}\lambda\,  e^{-\int dt\,  {\cal L}(\{\lambda_i(t)\})}, \qquad  \overline{\cal Q}(\{\lambda_i\},\{\mu_i\}) \equiv {\cal D}\lambda {\cal D}\mu \, e^{-\int dt\,  \overline{\cal L}(\{\lambda_i\},\{\mu_i\})} \EEQ
 may be interpreted as a space-time action.

The kernel $K$ of the quadratic form $(\mu,\mu)=\half \int dt \, dt'\, \sum_{i,j}
K_{ij}(t-t')\mu_i(t)\mu_j(t')$ appearing in the action is in this formalism equal to the covariance of the
noise in the original Langevin equation, here $\langle \dot{B}_i(t) \dot{B}_j(t')\rangle=K_{ij}(t-t')=2\del_{i,j} \del(t-t').$

\medskip


\subsection{Equilibrium Virasoro constraints: a reminder}


The so-called loop (or Virasoro) constraints are a well-known invariance statement
in the equilibrium theory which has proved extremely useful in obtaining
formulas in various asymptotic regimes for $N\to\infty$ (see e.g. \cite{DiFGinZJ} or
\cite{Joh}). We prove them using some elementary integration-by-parts trick on the
equilibrium measure.
Letting $\langle \ \cdot\ \rangle=
{\cal P}_{eq}(\, \cdot\, )$,
\BEQ 0=\sum_i \int d\lambda_i\, \frac{\partial}{\partial\lambda_i}\left(
 F(\{\lambda_j\}) e^{-W(\{\lambda_j\})} \right) = \sum_i
\left( \langle \frac{\partial F}{\partial \lambda_i} \rangle - \langle F
\frac{\partial W}{\partial \lambda_i} \rangle \right) \label{eq:eq-IPP} \EEQ
for any function $F$. 
Using this identity we can rederive very easily the well-known equilibrium Virasoro constraints (see e.g. \cite{DiFGinZJ} for a physical approach)  in
their general form given in Adler-Van Moerbeke
\cite{AvM}, restricting to $\beta=2$. Consider a potential $V_0(\lambda)$ such that 
$V'_0(\lambda)\equiv\sum_{k\ge 1} b_k\lambda^k$, where  all but a finite number of the coefficients $(b_k)_{k\ge 0}$ are zero. Perturb it formally by letting $V(\lambda)\equiv V[\tau](\lambda):=V_0(\lambda)+\sum_{k=1}^{+\infty} \tau_k \lambda^k$ depend on a set of parameters $\tau=\{\tau_k\}_{k\ge 0}$, and
write accordingly $Z_N(V)=Z[\tau]$, ${\cal P}={\cal P}[\tau]$, ${\cal P}_{eq}={\cal P}_{eq}[\tau]$, ${\cal Q}={\cal Q}[\tau]$, $\overline{\cal Q}=\overline{\cal Q}[\tau]$. Introduce the notation
\BEQ \pi_k \equiv \sum_i \lambda_i^k, \qquad k\ge 0 \label{eq:pi} \EEQ
for the sums of powers of the eigenvalues. Take $F(\{\lambda_i\})=\sum_i   \lambda_i^{n+1} $ in (\ref{eq:eq-IPP}). Then
\BEA  \langle \sum_i  (n+1)\lambda_i^{n}  \rangle &=&  
(n+1)  \langle \pi_{n} \rangle \nonumber\\
&=& \langle \lambda_i^{n+1}   \frac{\partial W}{\partial \lambda_i} \rangle  = \sum_{k=0}^{+\infty}  
b_k \langle \pi_{k+n+1} \rangle + \sum_{k=0}^{+\infty} k\tau_k \langle
\pi_{k+n}\rangle  -\beta  \sum_{i,j\not=i} \langle \frac{\lambda_i^{n+1}}{\lambda_i-\lambda_j}\rangle  \nonumber\\
&=& \sum_{k=0}^{+\infty} b_k \langle \pi_{k+n+1} \rangle + \sum_{k=0}^{+\infty}
k \tau_k \langle \pi_{k+n}\rangle -\frac{\beta}{2} \sum_{k=0}^{n}
\left( \langle \pi_k \pi_{n-k} \rangle -\langle \pi_{n}\rangle \right).  \nonumber\\
\label{eq:VC0} \EEA

Introduce the Fock representation of the oscillator algebra \cite{KacRai}
\BEQ \hat{a}_n=\beta^{1/2} \frac{\partial}{\partial\tau_n} \ (n\ge 1), \qquad 0  (n=0), \qquad  \beta^{-1/2} |n|\tau_{|n|} \ (n\le -1) \EEQ
and the associated free boson $\hat{a}(z)$ and energy-momentum tensor $\hat{L}(z)$,
\BEQ \hat{a}(z)=\sum_{n\in\Z} \hat{a}_n z^{-n-1}, \qquad \hat{L}(z)=\half
:(\hat{a}(z))^2: = \sum_{n\in\Z} \hat{L}_n z^{-n-2}. \EEQ
Noting that  
\BEQ \hat{L}_n= \sum_{k=0}^{+\infty} k\tau_k \frac{\partial}{\partial\tau_{k+n}} +
\frac{\beta}{2} \sum_{k=0}^n \frac{\partial}{\partial\tau_k} \frac{\partial}{\partial\tau_{n-k}},
\qquad n\ge -1 \EEQ
we see that (\ref{eq:VC0}) amouts to
\BEQ L_n^{eq}Z[\tau]=0 \EEQ
with 
\BEQ  L_n^{eq}=  \hat{L}_{n} + \beta^{-1/2} \left[ \sum_{k=0}^{+\infty} b_k \hat{a}_{n+k+1} + (\frac{\beta}{2}-1)(n+1)\hat{a}_n \right].
\label{eq:Vir-constraint} \EEQ

This two-line derivation has its interest, but the spirit of these constraints is really
of geometric origin, see e.g. \cite{AvM} or \cite{MirMor}: they reflect the way that the potential $V$ is transformed under
generators of conformal transformations $L_n=-\lambda^{n+1}\partial_{\lambda}$.

 
\subsection{Results of the article}


The aim of the article is to prove the existence {\em dynamical constraints} in the same spirit as the equilibrium Virasoro constraints discussed in the previous
subsection.

\medskip\noindent As pointed out just above, the conventional way to prove Virasoro constraints, see e.g. \cite{AvM} or \cite{MirMor},  is
 to consider the transformation of the equilibrium measure under a 
 conformal transformation  of the eigenvalues, $\lambda\mapsto \lambda+\eps \lambda^{n+1}$. 

\noindent In the dynamical case we miss a straightforward analogue of  (i) conformal transformations; (ii) the equilibrium
measure. Let us discuss these two points.

\begin{itemize}
\item[(i)]
 Our first claim is the following. The analogue of the group of conformal transformations
in the dynamical case is the group of {\em noise-preserving transformations}, briefly introduced in section 1 and discussed in full details in section \ref{def:higher-order-extension}, see in particular Definition \ref{def:noise-preserving-transformations} for the Lie
algebra of this group.  The corresponding infinitesimal transformations are "causality-preserving" transformations of the set of trajectories $\{\lambda(t),t\ge 0\}$, with
a condition called {\em noise-invariance condition}, see (\ref{eq:z=2}) 
or (\ref{eq:z=2bis}), ensuring that these
preserve the strength of the noise for trajectories satisfying a Langevin equation. 
\smallskip
This group contains in particular as a subgroup the {\em Schr\"odinger-Virasoro group}, an infinite-dimensional
group of coordinate transformations studied in details in the book \cite{RogUnt}, see
also \cite{Hen}. Briefly said, these are coupled space- and time-transformations
which are affine in space, thus defining an infinite-dimensional extension  of the two Virasoro generators $L_{-1,0}$. In 1D the Lie algebra is
generated by 
\BEQ X_f:=-f(t)\partial_t-\half \dot{f}(t) \lambda\partial_{\lambda}, \qquad 
Y_g:=-g(t)\partial_{\lambda}.\EEQ
While $(Y_g)$ is simply the time-current generated by $L_{-1}$, the $(X_f)$ are
local space-time transformations generalizing the infinitesimal scaling transformation
 $-t\partial_t-\half \lambda\partial_{\lambda}$ with dynamical exponent $z=2$, which  generates the  parabolic scaling 
 transformation  $(t,\lambda)\mapsto (a^2 t,a\lambda)$. This scaling originates from the
 transformation properties of white noise.  Noise-preserving infinitesimal transformations {\em not} belonging to the
Schr\"odinger-Virasoro algebra may be seen as the sum of a very general transformation of the
$\lambda$-coordinate, $\lambda\mapsto \lambda+\eps \del\lambda$, with
$\del\lambda(t)=f(\lambda(t))\Phi(t,\lambda)$, where $\Phi(t,\cdot)$ is some
{\em time-integrated functional} of the {\em past} of the trajectory,  and of a {\em time-transform} depending on $\lambda$, which suggests to
introduce the notion of a {\em proper time} (see section \ref{sec:space-time}). 
Though we are here in space dimension 1, the extension to $d$ space-dimensions is 
more or less straightforward;  using the conformal invariance of Brownian motion, it
is enough to require that the function $f$ in factor in the $\lambda$-coordinate
transform should define a conformal transformation.

\item[(ii)] Turning to the second point, it is not clear to us if there is a straightforward analogue of the
equilibrium measure. Naively, the partition function should be replaced by
the measure on trajectories, which is automatically normalized, and thus cannot be
used as a generating functional. However, perturbing the measure in the way of
Adler-Van Moerbeke (see previous subsection), one is led very naturally to an
un-normalized perturbed measure on the trajectories, ${\cal Q}^{lin}[\tau]={\cal Q}^{lin}[\tau](\{\lambda_i\})$ (see Definition \ref{def:generating-functional})  whose
integral ${\cal Z}^{lin}[\tau]:=\int d{\cal Q}^{lin}[\tau](\{\lambda_i\})$ may serve as
{\em generating functional}. The upper index "lin" stands for "linear", since ${\cal Q}[\tau]$ is obtained from the original measure on the trajectories by linearizing in
the $\tau$-parameters and then throwing away quadratic terms produced by the two-body
potentieal. 

Our main result is then Theorem \ref{th:main}, stating the {\em invariance of the
 generating functional ${\cal Z}^{lin}[\tau]$ under  Schr\"odinger-Virasoro 
transformations}. The action of these transformations is similar in aspect to the
action of Virasoro transformations on the partition in the equilibrium measure, see
(\ref{eq:Vir-constraint}), with the considerable difference though that we restricted
ourselves to indices $n=-1,0$, {\em but} on the other hand we have an infinite number
of constraints because of the arbitrary time-dependence. It exhibits a sum of  linear and of  quadratic expressions in
terms of a {\em static free boson} $\hat{\phi}(z,t)$ -- the free boson of usual conformal
field theory, with an extra, trivial time-dependence -- and of a {\em dynamical free boson} $\hat{\psi}(z,t)$ defined via a kernel $K$ depending on $V_0$ (see
Definition \ref{def:equation-motion}, Definition \ref{def:free-boson} and 
Definition \ref{def:free-boson-algebra}). The kernel $K=K(z^{-1},w)$, one of the
main ingredients in the computations, is the Green function of
the operator $D:=\partial_t+(\frac{\beta}{2}-1) \frac{d^2}{dz^2}-\frac{d}{dz}b(z)$ acting
on formal series $a_0+a_1 z+ a_2 z^2+\ldots$ 

Just as equilibrium Virasoro constraints may be used to compute the $n$-point functions
 of the first few so-called linear
statistics, $\pi_k:=\sum_{i=1}^N \lambda_i^k$, formula (\ref{eq:n-point-functions}), which
we reproduce here,  
\BEQ \Big{\langle} \left(\int dt\,  f_1(t)\pi_{k_1}(t)\right)\cdots 
 \left(\int dt\,  f_p(t)\pi_{k_p}(t)\right) \Big{\rangle}_0 =
 \prod_{q=1}^p  \left( -\int dt\, f_q(t) (K\ast \widehat{\partial/\partial\tau})_{k_q}(t) \right)
 {\cal Z}^{lin}[\tau]  \Big|_{\tau=0} \EEQ
shows that $n$-point functions may be obtained from ${\cal Z}^{lin}[\tau]$ by the
differentiation "trick" $\pi_k\equiv(K\ast \widehat{\partial/\partial\tau})_{k}(t)$
or (in terms of generating series) $\pi(z)\equiv -(K\ast\widehat{\partial/\partial\tau})(z,t)$. Note that if one had {\em not} linearized the generating functional,
we would have to solve instead a complex Burgers equation, $D\pi(z)+(\pi^2)'(z)=-
\widehat{\partial/\partial\tau}(z).$

\end{itemize}


\subsection{Perspectives}


In a future article, we plan  to extend Schr\"odinger-Virasoro constraints
to a much more general class of constraints, one per generator of the Lie algebra
of noise-preserving transformations. Formulas in Theorem (\ref{th:main}) being
readily generalized to arbitrary $n\not=-1,0$, it seems very likely that some of 
the conformal field theoretic structure uncovered for $n=-1,0$ will survive. As can be
expected, such
an extension is however far from obvious. Preliminary computations (some of them
presented already in the computations of \S 4.3, which often hold for arbitrary $n$, and
also ) 
show  that the action of more general transformations on the 
two-body potential produces cubic terms, plus  an infinite-number of new terms due
to the particle-dependent time-shifts  with unresolved singularity, typically $\int_0^t ds \frac{\lambda_i(s)-\lambda_j(s)}{(\lambda_i(t)-\lambda_j(t))^2}$ 
$(i\not=j)$ (see \S 1.2 B), which are however amenable to analysis provided one retains only a finite
number of terms in some perturbative expansion, either short-time or large $N$, with
foreseeable applications to the study of the limit $N\to\infty$ in the
microscopic regime. 


\subsection{Outline of the article}


Section 1 is an appetizer for the reader willing to understand the objective of the
paper and to have a flavour of the computations. The {\em noise invariance condition}
is introduced  right from the beginning in (\ref{eq:z=2}), but we postpone the general
discussion of this condition and consider only {\em elementary transformations} such
as (\ref{eq:coordinate-change}), which do not close under Lie brackets. The
transformation of the force term under these transformations is given in (\ref{eq:delV'})
for $N=1$ and (\ref{eq:3.5},\ref{eq:del-delay}) for general $N$. The very
complicated term (\ref{eq:del-delay})  fortunately vanishes for Schr\"odinger-Virasoro
 transformations, for which all time shifts are equal.

\medskip
 
 \noindent We present the noise invariance condition in whole generality in the
strictly  algebraically-minded section 2
 and define the {\em Lie algebra of noise-preserving transformations} ${\cal F}_{NP}$ in
 Definition \ref{def:noise-preserving-transformations}. We also compute the Lie
 brackets of elementary transformations in the natural {\em basis of iterated integrals}.
 
 \medskip
\noindent The very short Section 3 lies a general geometric foundation to these sets of
transformations. It is also the occasion to introduce the Schr\"odinger-Virasoro transformations.

\medskip
\noindent The main section is Section 4.  We present the key ingredients in
\S 4.1: the generating functional (Definition \ref{def:generating-functional});
generating series for the linear statistics and the parameters; 
the equation of motion for the linear statistics (Definition \ref{def:equation-motion});
the algebra of static and dynamic free bosons (Definition \ref{def:free-boson})
and its commutators (Definition \ref{def:free-boson-algebra}. Then comes the
statement of our main result, Theorem \ref{th:main}, yielding dynamical constraints
parallel to the equilibrium constraints of \S 0.2. The rest of the section is
devoted to the proof of Theorem \ref{th:main}.

\medskip\noindent Finally, we collected some technical lemmas used in the proofs
in an appendix (section 5).



\section{A short computational introduction}


The purpose in this section is to introduce and motivate the fundamental
 {\em noise-invariance condition} in a simplified setting, and to show some
preliminary computations in the case $N=1$ and in the general case, paving the
way to the more involved computations of section \ref{sec:dynamical-constraints}.


\subsection{The case $N=1$}
 

For pedagogical reasons we start from the case $N=1$, a one-dimensional general
Langevin equation, 
\BEQ d\lambda_t=dB_t-V'(\lambda_t)dt. \label{eq:L-1}\EEQ
We look for  infinitesimal transformations of the set of trajectories,  
\BEQ \{\lambda(t),t\ge 0\}\mapsto \{\lambda(t)+\eps \ (\bar{\del}\lambda)(t),t\ge 0\} \EEQ
that preserve the general structure
of the equation. We actually restrict to causality-preserving, first-order transformation,
namely, we assume that 
\BEQ \eps(\bar{\del} \lambda)(t)=\eps\left(\phi(t,\lambda) - \psi(t,\lambda) \dot{\lambda}(t) \right)\EEQ
where $\phi(t,\lambda),\psi(t,\lambda)$ 
depend only on the values of $(\lambda_s)_{s\le t}$.
Since the trajectory $t\mapsto\lambda(t)$ is not differentiable, this should be 
understood (to order one in $\eps$) as the composition of two transformations,
\BEQ \lambda\mapsto \lambda+\eps\del\lambda, \qquad t\mapsto t+\eps\del t \EEQ 
where
\BEQ (\del\lambda)(t)=\phi(t,\lambda), \qquad \del t= \psi(t,\lambda).\EEQ

 Put in another way, we look for the dynamical law satisfied by the transformed trajectory $\tilde{\lambda}(t+\eps\del t):=(\lambda+\eps\del\lambda)(t)$, or (to order 1 in $\eps$)
$\tilde{\lambda}(t)=(\lambda+\eps\del\lambda)(t-\eps\del t)$.  Since $dB_{t-\eps\del t}
=(1-\eps (\dot{\del t}))^{1/2} d\tilde{B}_t$ where $\tilde{B}$ has the same law as $B$, we get to
order 1 in $\eps$, taking 
into account the It\^o correction written as "Ito" in the following formula,
\BEQ d\tilde{\lambda}=\eps\left(\frac{\partial(\del \lambda)}{\partial t}+ {\mathrm{Ito}}
\right) dt+ (1+\eps\frac{\partial(\del\lambda)}{\partial\lambda}) \left\{ -(1-\eps (\dot{\del t})) V'(\lambda) dt+
(1-\frac{\eps}{2}(\dot{\del t})) d\tilde{B} \right\}. \EEQ

Under the fundamental {\em noise invariance condition}
\BEQ \frac{\partial(\del \lambda)}{\partial\lambda}=\half
(\dot{\del t}) \label{eq:z=2}. \EEQ
 (\ref{eq:L-1}) is turned into a similar Langevin equation with transformed force 
 $-(V'+\eps \del V')$ defined to order one in $\eps$ by 
\BEA && (V'+\eps\del V')(\lambda+\eps\del \lambda) = -\eps\frac{\partial (\del \lambda)}{\partial t}
+ \left[1+ \eps\left(\frac{\partial(\del \lambda)}{\partial\lambda} -(\dot{\del t}) \right) \right] V'(\lambda) -  \eps\  {\mathrm{Ito}} \nonumber\\
&& \qquad =  V'(\lambda+\eps\del \lambda)- \eps\left\{ \frac{\partial (\del \lambda)}{\partial t}
+ \frac{\partial(\del \lambda)}{\partial\lambda}
 V'(\lambda+\eps\del \lambda)+ V''(\lambda+\eps \del\lambda) \del\lambda +  {\mathrm{Ito}} \right\} \nonumber\\
 &&\qquad = V'(\lambda+\eps\del \lambda)- \eps\left\{ \frac{\partial (\del \lambda)}{\partial t}
+ \frac{\partial}{\partial\lambda}\left( \del \lambda\,   
 V'(\lambda+\eps \del \lambda)\right) + {\mathrm{Ito}}\right\} \nonumber\\
\EEA

\bigskip

\noindent Looking for specific examples, we now specialize  to the
 transformations where $\phi(t,\lambda)$ depends only on the value of $\phi$
 at time $t$,  namely (for $n\ge -1$)
\BEQ \del\lambda(t)=- \lambda^{n+1}(t) \dot{a}(t), \ \ \del t=2\int_0^t ds\, \frac{\partial(\del \lambda)}{\partial\lambda}(s) = -2(n+1) \int_0^t ds\, \dot{a}(s) \lambda^{n} (s),
\label{eq:coordinate-change} \EEQ
For $n=-1,0$ these infinitesimal trajectory transformations may be seen as simple {\em coordinate transformations}; they generate the {\em Schr\"odinger-Virasoro algebra} introduced in
section \ref{sec:space-time}. For $n\ge 1$ however, these transformations act on the whole
trajectory, and commutators  generate a much larger class of transformations studied
in the next section. 

\smallskip

\noindent With  an It\^o correction
${\mathrm{Ito}}=\frac{\partial^2(\del\lambda(t))}{\partial\lambda(t)^2}=-(n+1)n\lambda^{n-1}(t) \dot{a}(t)$ in this specific case, we get our first important formula,

\vskip 1cm

{\bf{ \qquad  ($N=1$ force change)}}
\BEQ \del V'=\left(\lambda^{n+1} \ddot{a}+ \left\{ \sum_{k\ge 0}
b_k (n+1+k)
\lambda^{n+k}  + (n+1)n\lambda^{n-1}  \right\} \dot{a} \right). \label{eq:delV'} \EEQ

\bigskip

All these terms extend trivially to the case of $N$ particles when $\beta=0$, i.e. in
absence of two-body potential, yielding $N$ terms, $\del V'_i$, where
\BEQ V'_i:=-\frac{\partial W}{\partial\lambda_i}= \sum_{j\not=i} \frac{\beta}{\lambda_i-\lambda_j}-V'(\lambda_i)  \label{eq:V'i} \EEQ
is the {\em force felt by the $i$-th particle}.
The first term in (\ref{eq:delV'}) reflects the time-dependence of the transformation.
The last term is the It\^o's correction. The second term expresses simply the
action  of the Virasoro vector field $-\dot{a}(\lambda^{n+1}\frac{\partial}{\partial\lambda}+(n+1)
\lambda^n)$ on the confining force $-V'$.


\subsection{The $N$-particle model}


General transformations leaving invariant the form of the equation for $N\ge 1$
will imply different time-changes for the $N$ particles located at $\{\lambda_i\}$,
as is immediately seen from the noise invariance condition (\ref{eq:z=2}); see section 
\ref{sec:space-time}
for general geometric considerations.  This makes in general the transformation of the two-body force more complicated, though (as we shall see later on) the
change in the action remains surprisingly simple. 

Generally speaking the {\em change of the force felt by the $i$-th particle} (see
(\ref{eq:V'i})) or
simply {\em force change}, $\del V'_i$,  is the sum of two terms. The first one
(thereafter called {\em simultaneous}), $\del_{simul}V'_i$, is the more or less straightforward of the
$N=1$ force change written in the previous subsection, taking also into account the
action of the coordinate change on the two-body force. The second (called {\em
delayed}), $\del_{delay}V'_i$,  takes into
account the difference of time-shifts between two trajectories $(\lambda_i(t))_{t\ge 0}$
and $(\lambda_j(t))_{t\ge 0}$, $i\not=j$.

\bigskip

\noindent{\bf A. Simultaneous force change}

\medskip 

 Compared with
the $N=1$ case,  we must now write down the effect on the dynamics of $\lambda_i$ of the
coordinate change. In addition to the term (\ref{eq:coordinate-change}), one has an extra term  due to the transformation 
of the two-body force, which must also take into account the transformation of the other
eigenvalues $\{\lambda_j\}_{j\not=i}$,   
\BEA &&   -\dot{a} \left( \sum_{i'=1}^N \lambda^{n+1}_{i'} \frac{\partial}{\partial\lambda_{i'}}
+(n+1)\lambda_i^n \right)  \left(  \sum_{j\not=i} \frac{\beta}{\lambda_i-\lambda_j}  \right)= -\beta
 \left\{
 \sum_{j\not=i}  \frac{\lambda_i^{n+1} -\lambda_j^{n+1} }{(\lambda_i-\lambda_j)^2} 
\right.  \nonumber\\ && \qquad \left. -
 (n+1) \sum_{j\not=i} \frac{ \lambda_i^n}{\lambda_i-\lambda_j}  \right\} \dot{a}(t)
  \equiv  - \left[ \sum_{k=0}^{n}  A_{n,k} \right] \dot{a}(t), \EEA
 where  
\BEQ A_{n,k}:=\beta \sum_{j\not=i} 
 \frac{\lambda_i^k \lambda_j^{n-k} - \lambda_i^n}{\lambda_i-\lambda_j}
 =\beta \lambda_i^k  \sum_{p=0}^{n-1-k} \lambda_i^p \sum_{j\not=i}  \lambda_j^{n-1-k-p},\EEQ
 from which 
 \BEQ \sum_{k=0}^n A_{n,k}= \beta \sum_{q=0}^{n-1} 
   (q+1) \lambda_i^q \sum_{j\not=i}\lambda_j^{n-1-q}= \beta \left[  \sum_{q=0}^{n-1}
(q+1) \lambda_i^q \pi_{n-1-q} - \half (n+1)n \lambda_i^{n-1} \right]. 
 \label{eq:3.3}\EEQ 
 
Changing sign, we may interpret (\ref{eq:3.3}) as an additive contribution to 
$\del V'_i$. Combining with the It\^o term, see third term in (\ref{eq:coordinate-change}),
 we get 
\BEQ   \sum_{k=0}^{n} A_{n,k}+ (n+1)n \lambda_i^{n-1}
 = \beta\sum_{q=0}^{n-1} 
(q+1) \lambda_i^q \pi_{n-1-q}
 +(1-\frac{\beta}{2}) (n+1)n\lambda_i^{n-1} . \EEQ

\medskip

\noindent The other terms in the action transform as in section 1 (compare with
(\ref{eq:coordinate-change})), yielding a total variation

\bigskip

\qquad \qquad {\bf{(simultaneous force change for general N)}}

\BEA &&  \del_{simul} V'_i= \lambda_i^{n+1} \ddot{a}+ \left\{ \sum_{l=0}^{+\infty} b_l (n+l+1)\lambda_i^{n+l} \nonumber \right. \\ &&  \qquad \left.  +   \left[ \beta \sum_{q=0}^{n-1} 
 (q+1)\lambda_i^q \pi_{n-1-q}
 -         (\frac{\beta}{2}-1) (n+1)n\lambda_i^{n-1} 
  \right]\,  \right\} \ \dot{a}. \nonumber\\
 \label{eq:3.5}  \EEA

We return to  these computations in section \ref{sec:dynamical-constraints} after a 
more detailed discussion of the noise-preserving condition.

\bigskip

\noindent {\bf{B. Delayed force change}}

\medskip

Consider only the part of the variation $\bar{\del}\lambda$ due to the time-shifts,
\BEQ \del t_i:=2(n+1)\int_0^t ds\, \dot{a}(s)\lambda_i^n(s).\EEQ
Letting $\tilde{\lambda}_i(t):=\lambda_i(t-\eps\del t_i)$, the system of coupled
equations for the particles becomes (to first order in $\eps$)
\BEA &&  d\tilde{\lambda}_i(t)=\left(1-\eps (n+1) \dot{a}(t) \tilde{\lambda}^n_i(t) \right)
\, d\tilde{B}_i(t) - \left(1-2\eps(n+1) \dot{a}(t) \tilde{\lambda}_i^n(t)\right)
V'(\tilde{\lambda}_i(t)) \, dt \nonumber\\
&& \qquad\qquad + \left(1-2\eps(n+1) \dot{a}(t) \tilde{\lambda}_i^n(t) \right)
\sum_{j\not=i} \frac{\beta}{\tilde{\lambda}_i(t)-\lambda_j(t-\eps\del t_i)} \, dt
\label{eq:delayed-force} \EEA
where  $\frac{1}{\sqrt{2}}(\tilde{B}_i)_i$ are standard Brownian motions. Adding
to this variation the one due to $\del\lambda$ compensates the change of noise strength
due to the noise invariance condition. The last term in the r.-h.s. of (\ref{eq:delayed-force}) brings to light a new effect due to the different time-shift. Since
$\lambda_j(t-\eps\del t_i)=\tilde{\lambda}_j(t+\eps(\del t_j-\del t_i))$, we have to
order $1$ in $\eps$
 \BEQ \frac{\beta}{\tilde{\lambda}_i(t)-\lambda_j(t-\eps\del t_i)}= \frac{\beta}{\tilde{\lambda}_i(t)-\tilde{\lambda}_j(t)} + \eps \ \frac{\beta}{(\tilde{\lambda}_i(t)-
 \tilde{\lambda}_j(t))^2} \, \cdot\, \frac{d\tilde{\lambda}_j(t)}{dt} (\del t_j-\del t_i) \EEQ
Thus (combining with the effect of the $\del\lambda$-variation studied in A.),
 $\lambda'_i=\lambda_i+(\bar{\del}\lambda)_i$, $i=1,\ldots,N$ follow the modified
system of equations to first order in $\eps$
\BEQ  \frac{d\lambda'_i}{dt} = d\tilde{B}_i(t) -
 (V'_i(t,\lambda')+\eps \del_{simul}V'_i(t,\lambda')) dt + \eps
\sum_{j\not=i} (\del t_j-\del t_i) \frac{\beta}{(\lambda'_i(t)-\lambda'_j(t))^2}
\ \frac{d\lambda'_j}{dt} \label{eq:1.21}.\EEQ
Replacing $\frac{d\lambda'_j}{dt}$ by the $0$-th order term $d\tilde{B}_j(t)-\frac{\partial W}{\partial\lambda'_j}(t)dt$
in the right-hand side of (\ref{eq:1.21}), we get
\BEQ \frac{d\lambda'_i}{dt}=dB'_i(t) -(V'_i(t,\lambda')+\eps \del_{simul}V'_i(t,\lambda'))dt - \eps\ \beta \sum_{j\not=i} \frac{\partial W}{\partial\lambda'_j}(t) \frac{\del t_j-\del t_i}{(\lambda'_i(t)-\lambda'_j(t))^2} \, dt,\EEQ
where 
\BEQ dB'_i(t):=d\tilde{B}_i(t)+\eps \sum_{j\not=i} (\del t_j-\del t_i)
\frac{\beta}{(\lambda'_i(t)-\lambda'_j(t))^2} \, d\tilde{B}_j(t), \qquad i=1,\ldots,n \EEQ
have same law as the original Brownians since the $\eps$-term defines an infinitesimal
rotation and white noise is invariant by rotation.  Thus we have found
\BEQ \del_{delay}V'_i(t)=-  \beta \sum_{j\not=i} \frac{\partial W}{\partial\lambda'_j}(t) \frac{\del t_j-\del t_i}{(\lambda'_i(t)-\lambda'_j(t))^2}. \label{eq:del-delay} \EEQ

\bigskip
{\bf{C. Change of measure}}

\medskip
Let us finally discuss the change of measure on the trajectories -- we shall return to this in section
\ref{sec:dynamical-constraints} with a modified, $\tau$-dependent measure.

Comparing the measure ${\cal Q}(V')$, resp. ${\cal Q}(V'+\del V')\equiv {\cal Q}(V')+
\del {\cal Q}$ on the space of trajectories of (\ref{eq:L}) with confining forces
$\{-V'(\lambda_i)\}_i$, resp. $\{-(V'(\lambda_i)+\del V'_i)\}_i$, we see from (\ref{eq:Q})
or rather from the rigorous Girsanov formula (\ref{eq:Girsanov}) that
\BEA  \del {\cal Q}(\{\lambda_i\}) &=& {\cal Q}(\{\lambda_i\})  \sum_i \int \del V'_i \, dB_i(t)\nonumber\\ &=&  {\cal Q}(\{\lambda_i\})\  \sum_i \Big( \int \del V'_i \,  d\lambda_i(t)+ \int \del V'_i\  \frac{\partial W}{\partial \lambda_i}(t)\, dt \Big). 
\label{eq:change-of-measure} \EEA

The main technical task in section 4 is to compute the terms appearing in (\ref{eq:change-of-measure}) in the case of  Schr\"odinger-Virasoro transformations, for which  $\del V'_i=\del_{simul}V'_i$ simply.


\section{Higher-order extension}


We shall now construct the Lie algebra generated by the transformations (\ref{eq:coordinate-change}). 

\begin{Definition} \label{def:higher-order-extension}
Let ${\cal F}$ be the space of functionals $\Phi=\Phi(t,\lambda)$ generated (as
as vector space) by functionals of the form
\BEQ \dot{a}(t)\int_0^t ds_1 \, \dot{a}_1(s_1) \lambda^{k_1}(s_1) \int_0^{s_1}
ds_2\, \dot{a}_2(s_2) \lambda^{k_2}(s_2)\cdots \int_0^{s_{p-1}}ds_p\ \dot{a}_p(s_p)
\lambda^{k_p}(s_p) \qquad (p\ge 0, \ k_1,\ldots,k_p\ge 0) \label{eq:higher-order-extension}\EEQ
where $\dot{a},\dot{a}_1,\ldots,\dot{a}_p$ are smooth functions of time.
\end{Definition}

Integrals 
\BEQ \Phi^{(k_1,\ldots,k_p)}(\dot{a}_1,\ldots,\dot{a}_p;\lambda)(t):=\int_0^t ds_1\, \dot{a}_1(s_1) \lambda^{k_1}(s_1) \int_0^{s_1} ds_2\, \dot{a}_2(s_2) \lambda^{k_2}(s_2)\cdots\int_0^{s_{p-1}}ds_p \, \dot{a}_p(s_p)\lambda^{k_p}(s_p) 
\EEQ
 are called {\em iterated integrals}. As a prominent example, a completely factorized functional
$\prod_{i=1}^p \left(\int_0^t ds_i \, \dot{a}_i(s_i) \lambda^{k_i}(s_i) \right)$ is
a sum of $p!$ iterated integrals since (denoting by $\Sigma_p$ the group of permutations
of a set of $p$ elements)
\BEQ \int_0^t ds_1 \int_0^t ds_2\cdots \int_0^t ds_p
\left(\cdots\right)=\sum_{\sigma\in\Sigma_p} \int_0^t ds_{\sigma(1)} \int_0^{s_{\sigma(1)}} ds_{\sigma(2)} \cdots \int_0^{s_{\sigma(p-1)}} ds_{\sigma(p)} \left(\cdots\right). \EEQ
The class ${\cal F}$ is stable by multiplication because of the shuffle relation,
\BEQ \left[ \int_0^t ds_1 \int_0^t ds_2\cdots \int_0^t ds_p
\left(\cdots\right) \right] \left[ \int_0^t ds_{\bar{1}} \int_0^t ds_{\bar{2}}\cdots \int_0^t ds_{\bar{q}}
\left(\cdots\right) \right] = \nonumber \EEQ
\BEQ \qquad =  \sum_{\sigma} \left[ \int_0^t ds_{\sigma(1)} \int_0^t ds_{\sigma(2)}\cdots \int_0^t ds_{\sigma(p+q)}
\left(\cdots\right) \right]  \EEQ
where $\sigma$ ranges over shuffles of the lists $(1,\ldots,p)$, $(\bar{1},\ldots,\bar{q})$, i.e.
over all re-orderings of the compound list $(1,\ldots,p,\bar{1},\ldots,\bar{q})$  preserving the orderings of the two sub-lists. In particular, the prefactor $\dot{a}(t)$ in (\ref{eq:higher-order-extension}) may be interpreted as a multiplication by
$\int_0^t ds_0 \, \ddot{a}(s_0)$ and absorbed into an iterated integral of order $p+1$. Finally, "polarizing" a $p$-th iterated integral by
replacing $\lambda$ with $p$ independent copies $\lambda_1,\ldots,\lambda_p$, namely,
\BEQ 
\Phi^{(k_1,\ldots,k_p)}(\dot{a}_1,\ldots,\dot{a}_p;\lambda_1,\ldots,\lambda_p)(t):=\int_0^t ds_1\, \dot{a}_1(s_1) \lambda_1^{k_1}(s_1) \int_0^{s_1} ds_2\, \dot{a}_2(s_2) \lambda_2^{k_2}(s_2)\cdots\int_0^{s_{p-1}}ds_p \, \dot{a}_p(s_p)\lambda_p^{k_p}(s_p)  
\EEQ
and permuting the order of integration by use of Fubini's theorem, we obtain after some
computations (see \cite{Unt1} or \cite{Unt2})
\BEA && 
\Phi^{(k_1,\ldots,k_p)}(\dot{a}_1,\ldots,\dot{a}_p;\lambda_1,\ldots,\lambda_p)(t)= 
\nonumber\\
&& \qquad =
\sum_{\sigma\in\Sigma_p} \eps(\sigma) \int_0^t ds_1\, \dot{a}_{\sigma(1)}(s_1) \lambda_{\sigma(1)}^{k_{\sigma(1)}}(s_1) \int_0^{s_1} ds_2\, \dot{a}_{\sigma(2)}(s_2) \lambda_{\sigma(2)}^{k_{\sigma(2)}}(s_2)\cdots\int_0^{s_{p-1}}ds_p \, \dot{a}_{\sigma(p)}(s_p)\lambda_{\sigma(p)}^{k_{\sigma(p)}}(s_p)   \nonumber\\
\EEA
for some universal coefficients $\eps(\sigma)\in\Z$. Alternatively, if 
$\Phi(\lambda)\equiv \Phi^{(k_1,\ldots,k_p)}(\dot{a}_1,\ldots,\dot{a}_p;\lambda)$, then we get  
\BEQ \Phi(\lambda_1,\ldots,\lambda_p)(t)=\sum_{\sigma\in\Sigma_p} \eps(\sigma)
\Phi^{\sigma}(\lambda_{\sigma(1)},\ldots,\lambda_{\sigma(p)})(t)\EEQ
by defining $\Phi^{\sigma}(\lambda_1,\ldots,\lambda_p)(t)$ to be the 
polarization of $\Phi^{\sigma}(\lambda)(t):=\Phi^{(k_{\sigma(1)},\ldots,k_{\sigma(p)})}(\dot{a}_{\sigma(1)},\ldots,\dot{a}_{\sigma(p)};\lambda)(t).$
This polarization trick will allow us later
on to evaluate 
$\frac{d}{d\eps}\big|_{\eps=0} \Phi(\lambda+\eps\bar{\del}\lambda)(t)$ as the sum 
\BEQ \sum_{\sigma\in\Sigma_p} \eps(\sigma)
\frac{d}{d\eps}\big|_{\eps=0} \Phi^{\sigma}(\lambda+\eps\bar{\del}\lambda,\lambda,\ldots,
\lambda)(t).  \label{eq:polarization-trick} \EEQ

Note that, by completing the tensor product, we may also choose 
to replace $\dot{a}_1(s_1)\dot{a}_2(s_2)\cdots
\dot{a}_p(s_p)$ with a general time coefficient $g(s_1,\ldots,s_p)$ in Definition
\ref{def:higher-order-extension}.

\bigskip
\noindent Now comes our main definition.

\begin{Definition}[noise-preserving transformations] \label{def:noise-preserving-transformations}
Let ${\cal F}_{NP}$ be the Lie algebra generated (as a vector space) by  infinitesimal trajectory transformations of the type
\BEQ (\bar{\del}\lambda)(t)=\lambda(t)^{n+1} \Phi(t,\lambda) - \dot{\lambda}(t) \Psi(t,\lambda) \label{eq:bar-del} \EEQ
with $\Phi\in{\cal F}$ and 
\BEQ \Psi(t,\lambda)=2(n+1) \int_0^t ds\, \lambda(s)^n \Phi(s,\lambda). \label{eq:z=2bis} \EEQ
\end{Definition}

Replacing as in the previous section the infinitesimal transformation 
$\lambda\mapsto \lambda+\eps \bar{\del}\lambda$ by the composition of 
$\lambda\mapsto \lambda+\eps\del\lambda=\lambda+\eps \lambda^{n+1}\Phi(\cdot,\lambda)$
with the time-transformation $t\mapsto t+\eps \Psi(t,\lambda)$, we see that
(\ref{eq:z=2bis}) generalizes (\ref{eq:z=2}) in an obvious way to transformations depending on the past of the trajectory. For the sequel we note that:
\BEQ \half\partial_t \Psi(t,\lambda)=(n+1) \lambda(t)^n \Phi(t,\lambda)=
\frac{\partial}{\partial\lambda(t)} \Phi(t,\lambda), \EEQ
 where the {\em partial derivative} $\frac{\partial}{\partial\lambda(t)}$ (to be distinguished from the {\em functional
 derivative}
 $\frac{\del}{\del\lambda(t)}$) acts on the function $\lambda(t)^{n+1}$ but {\em vanishes}
 on the integrated functional $\Phi(t,\lambda)$.

\medskip

We want to compute the Lie bracket of two noise-preserving transformations 
(\ref{eq:bar-del}) and check that it is still a noise-preserving transformation.
We start by specializing to the case when
\BEQ (\bar{\del}_i\lambda)(t)=\lambda(t)^{n_i+1} \dot{a}_i(t) -2(n_i+1)
\dot{\lambda}(t) \int_0^t ds \, \dot{a}_i(s) \lambda^{n_i}(s), \qquad i=1,2.
\label{eq:elementary-transformation} \EEQ
Then 
\BEA && \left([\bar{\del}_1,\bar{\del}_2]\lambda\right)(t) =  \frac{\partial^2}{\partial
\eps_1 \partial\eps_2}\big|_{\eps_1=\eps_2=0}  \left[(\lambda+\eps_1\bar{\del}_1\lambda)+\eps_2\bar{\del}_2
(\lambda+\bar{\del}_1\lambda) - (\lambda+\eps_2\bar{\del}_2\lambda)-\eps_1\bar{\del}_1
(\lambda+\bar{\del}_2\lambda) \right] \nonumber\\
&& \qquad = \frac{\partial^2}{\partial
\eps_1 \partial\eps_2}\big|_{\eps_1=\eps_2=0} \left\{ 
 \eps_2\left[ \left(\lambda(t)+\eps_1 \lambda(t)^{n_1+1} \dot{a}_1(t) - 2\eps_1
 (n_1+1) \dot{\lambda}(t) \int_0^t ds\, \dot{a}_1(s) \lambda^{n_1}(s) \right)^{n_2+1}
 \dot{a}_2(t)  \right.\right.\nonumber\\
&& \left. \left. \qquad -2(n_2+1)\partial_t \left( \lambda(t)+\eps_1 \lambda(t)^{n_1+1} \dot{a}_1(t) - 2\eps_1
 (n_1+1) \dot{\lambda}(t) \int_0^t ds\, \dot{a}_1(s) \lambda^{n_1}(s) \right) \ \cdot\ 
 \right.\right.\nonumber\\
 && \left.\left. \qquad  \cdot\ \int_0^t ds\, \dot{a}_2(s) \left(\lambda(s)+\eps_1 \lambda(s)^{n_1+1} \dot{a}_1(s) - 2\eps_1
 (n_1+1) \dot{\lambda}(s) \int_0^s ds'\, \dot{a}_1(s') \lambda^{n_1}(s') \right)^{n_2}
\right]\ -\ (1\leftrightarrow 2) \right\} \nonumber\\ 
\label{eq:3.11} \EEA

Easy computations give $\left([\bar{\del}_1,\bar{\del}_2]\lambda\right)(t)=
\left[ F_I+F_{II}+ \left(F_1+F_2+F_3+F\right) \dot{\lambda}(t) \right](n_1,\dot{a}_1;
n_2,\dot{a}_2)- \left[ F_I+F_{II}+ \left(F_1+F_2+F_3+F\right) \dot{\lambda}(t) \right]
(n_2,\dot{a}_2;n_1,\dot{a}_1) $, with (abbreviating $F_{\cdot}(n_1,\dot{a}_1;n_2,\dot{a}_2)$ to $F_{\cdot}$):
\BEQ F_I:=\dot{a}_2(n_2+1) \lambda(t)^{n_1+n_2+1} \dot{a}_1(t);\EEQ
\BEQ F_{II}:=-2(n_2+1) \lambda(t)^{n_1+1} \ddot{a}_1(t) \int_0^t ds\, \dot{a}_2(s)
\lambda(s)^{n_2}; \EEQ
\BEQ F_1:=-2\dot{a}_2(t) (n_2+1)(n_1+1) \lambda(t)^{n_2} \int^t_0 ds\, \dot{a}_1(s)
\lambda(s)^{n_1};\EEQ
\BEQ F_{2}:=2(n_2+1)(n_1+1) \left(\int^t_0 ds\, \dot{a}_2(s) \lambda(s)^{n_2} \right)
\lambda(t)^{n_1} \dot{a}_1(t);\EEQ 
\BEQ F_{3}:=-2(n_2+1)n_2 \int_0^t ds\, \dot{a}_2(s) \lambda(s)^{n_1+n_2} \dot{a}_1(s);\EEQ
and (integrating by parts)
\BEQ F:=4(n_2+1)n_2(n_1+1) \int_0^t ds\, \dot{\lambda}(s) \dot{a}_2(s) 
\lambda(s)^{n_2-1} \int_0^s ds'\, \dot{a}_1(s') \lambda(s')^{n_1} \equiv
4(n_1+1)(n_2+1) (F_4+F_5+F_6), \EEQ
with
\BEQ F_4:= \lambda(t)^{n_2} \dot{a}_2(t) \int_0^t ds'\, \dot{a}_1(s') \lambda(s')^{n_1},
\qquad F_5:=-\int^t_0 ds\, \lambda^{n_2}(s) \ddot{a}_2(s) \int_0^s ds'\, \dot{a}_1(s')
\lambda^{n_1}(s')\EEQ
\BEQ F_6:=-\int_0^t ds\,  \lambda^{n_2+n_1}(s) \dot{a}_2(s)\dot{a}_1(s) \EEQ

There is also a term $F^*$ in $\ddot{\lambda}$, but due to symmetry 
$F^*(n_1,\dot{a}_1;n_2,\dot{a}_2)-F^*(n_2,\dot{a}_2;n_1,\dot{a}_1)=0$. For the same
reason, the two $F_6$-terms cancel. Finally, one remarks that $F_1,F_4,F_2$ are proportional
and sum up to $0$, while
 $\half \partial_t F_5=\frac{\partial}{\partial\lambda(t)} F_{II}$
and $\half \partial_t F_3=\frac{\partial}{\partial\lambda(t)} F_{I}$.

Concluding, 
\BEA  \left([\bar{\del}_1,\bar{\del}_2]\lambda\right)(t)
&=& \left(\lambda(t)^{n_1+n_2+1}
\Phi_{[\del_1,\del_2]}^{12}(t,\lambda)-\dot{\lambda}(t) \Psi_{[\del_1,\del_2]}^{12}(t,\lambda) \right) \nonumber\\
&+& 
\left(\lambda(t)^{n_1+1}
\Phi_{[\del_1,\del_2]}^{2}(t,\lambda)-\dot{\lambda}(t) \Psi_{[\del_1,\del_2]}^{2}(t,\lambda) \right) -
\left(\lambda(t)^{n_2+1}
\Phi_{[\del_1,\del_2]}^{1}(t,\lambda)-\dot{\lambda}(t) \Psi_{[\del_1,\del_2]}^{1}(t,\lambda) \right) \nonumber\\
 \EEA
is a noise-preserving transformation, with
\BEQ \Phi_{[\del_1,\del_2]}^{12}(t,\lambda)=(n_2-n_1) \dot{a}_2(t)\dot{a}_1(t); \EEQ
\BEQ  
\Phi_{[\del_1,\del_2]}^2(t,\lambda)=-2(n_2+1)\ddot{a}_1(t) \int_0^t ds\, \dot{a}_2(s) \lambda(s)^{n_2},
\qquad
\Phi_{[\del_1,\del_2]}^1(t,\lambda)=-2(n_1+1)\ddot{a}_2(t) \int_0^t ds\, \dot{a}_1(s) \lambda(s)^{n_1}
\EEQ

and $\Psi_{[\del_1,\del_2]}^{12}$, resp. $\Psi_{[\del_1,\del_2]}^2$, 
$\Psi_{[\del_1,\del_2]}^1$ associated to $\Phi_{[\del_1,\del_2]}^{12}$, resp. 
$\Phi_{[\del_1,\del_2]}^2$, 
$\Phi_{[\del_1,\del_2]}^1$ by (\ref{eq:z=2bis}).
The above formulas for $\Phi_{[\del_1,\del_2]}^2,\Phi_{[\del_1,\del_2]}^1$ show clearly the necessity to extend the set
of noise-preserving transformations by allowing iterated integrals.

\vskip 2cm

\noindent Consider now two general noise-preserving transformations $\bar{\del}_i$ with
\BEQ \bar{\del}_i\lambda(t)=\lambda(t)^{n_i+1} \Phi_i(t,\lambda)-2(n_i+1)\dot{\lambda}(t)
\int_0^t ds_1\, \lambda(s_1)^{n_i} \Phi_i(s_1,\lambda),\EEQ
with 
\BEQ \Phi_1(t,\lambda)=\int_0^t ds_2 \, \dot{a}_2(s_2) \lambda(s_2)^{k_2}\int_0^{s_2} ds_3\ \dot{a}_3(s_3)\lambda(s_3)^{k_3} \cdots\int_0^{s_{p-1}} ds_p\, \dot{a}_p(s_p)
\lambda(s_p)^{k_p}, \nonumber\\ \EEQ
\BEQ
\Phi_2(t,\lambda)=\int_0^t ds_2\, \dot{b}_2(s_2)\lambda(s_2)^{k'_2}\int_0^{s_2} ds_3\ \dot{b}_3(s_3)\lambda(s_3)^{k'_3} \cdots\int_0^{s_{p'-1}} ds_{p'}\, \dot{b}_{p'}(s_{p'})
\lambda(s_{p'})^{k_{p'}}. \EEQ
Let $\Phi_1^{\sigma}(\lambda_{\sigma(2)},\lambda\ldots,\lambda)(t)=:\int_0^t ds_2\,
 \dot{a}_{\sigma(2)}(s_2)
(\lambda_{\sigma(2)}(s_2))^{k_{\sigma(2)}}
B(\Phi_1^{\sigma})(s_2,\lambda)$ and similarly,\\
$\Phi_2^{\sigma}(\lambda_{\sigma(2)},\lambda,\ldots,\lambda)(t)=:\int_0^t ds_2\,
 \dot{b}_{\sigma(2)}(s_2)
(\lambda_{\sigma(2)}(s_2))^{k'_{\sigma(2)}}
B(\Phi_2^{\sigma})(s_2,\lambda)$.
Then we get 
\BEA && \left([\bar{\del}_1,\bar{\del}_2]\lambda\right)(t) = \left(\lambda(t)^{n_1+n_2+1}
\Phi_{[\del_1,\del_2]}^{12}(t,\lambda)-\dot{\lambda}(t) \Psi_{[\del_1,\del_2]}^{12}(t,\lambda) \right) \nonumber\\
&& \qquad + 
\left(\lambda(t)^{n_1+1}
\Phi_{[\del_1,\del_2]}^{2}(t,\lambda)-\dot{\lambda}(t) \Psi_{[\del_1,\del_2]}^{2}(t,\lambda) \right) -
\left(\lambda(t)^{n_2+1}
\Phi_{[\del_1,\del_2]}^{1}(t,\lambda)-\dot{\lambda}(t) \Psi_{[\del_1,\del_2]}^{1}(t,\lambda) \right) \nonumber\\
&&  +\ \  \left[ F_{I'}+F_{II'}+(F_{1'}+F_{2'})\dot{\lambda}(t)\right](n_1,\Phi_1;n_2,\Phi_2)-
\left[ F_{I'}+F_{II'}+(F_{1'}+F_{2'})\dot{\lambda}(t)\right](n_2,\Phi_2;n_1,\Phi_1) ,
 \nonumber\\ \EEA
where (substituting $\Phi_i(s,\lambda)$ to $\dot{a}_i(s)$ with respect to the
previous computations)
\BEQ \Phi_{[\del_1,\del_2]}^{12}(t,\lambda)=(n_2-n_1) \Phi_2(t,\lambda)\Phi_1(t,\lambda); \EEQ
\BEQ  
\Phi_{[\del_1,\del_2]}^2(t,\lambda)=-2(n_2+1)\partial_t\Phi_1(t,\lambda) \int_0^t ds\, \Phi_2(s,\lambda) \lambda(s)^{n_2},
\EEQ
\BEQ \qquad \qquad 
\Phi_{[\del_1,\del_2]}^1(t,\lambda)=-2(n_1+1)\partial_t\Phi_2(t,\lambda) \int_0^t ds\, \Phi_1(s,\lambda) \lambda(s)^{n_1}
\EEQ

and $F_{I'}(n_1,\Phi_1;n_2,\Phi_2)\equiv F_{I'},F_{II'}(n_1,\Phi_1;n_2,\Phi_2)\equiv
F_{II'},
F_{1'}(n_1,\Phi_1;n_2,\Phi_2)\equiv F_{1'},F_{2'}(n_1,\Phi_1;n_2,\Phi_2)\equiv
F_{2'}$ are new terms obtained by
letting the derivative $\frac{\partial}{\partial\eps_1}\big|_{\eps_1=0}$ act
on the $\lambda$-dependent terms $\Phi_2(t,\lambda), \Phi_2(s,\lambda)$ found instead
of $\dot{a}_2(t)$, resp. $\dot{a}_2(s)$ in the straightforward generalization of 
(\ref{eq:3.11}).  The polarization trick (\ref{eq:polarization-trick}) applied to
$\Phi_2$ yields
$F_{\cdot'}\equiv \sum_{\sigma\in\Sigma_{p-1}} \eps(\sigma) F_{\cdot'}(\sigma)$, with

\BEQ F_{I'}(\sigma)=\lambda(t)^{n_2+1} \int_0^t ds_2\, \dot{b}_{\sigma(2)}(s_2) k'_{\sigma(2)} \lambda(s_2)^{k'_{\sigma(2)}+n_1} \Phi_1(s_2,\lambda) B(\Phi_2^{\sigma})(s_2,\lambda)\EEQ
\BEQ F_{II'}(\sigma)=-2(n_1+1)\lambda(t)^{n_2+1} \int_0^t ds_2 \, \dot{b}_{\sigma(2)}(s_2)
 B(\Phi_2^{\sigma})(s_2,\lambda) k'_{\sigma(2)}\lambda(s_2)^{k'_{\sigma(2)}-1} \dot{\lambda}(s_2) \int_0^{s_2} ds_3\,
 \lambda(s_3)^{n_1} \Phi_1(s_3,\lambda)  \EEQ
 \BEQ  F_{1'}(\sigma)=-2(n_2+1) \int_0^t ds_1\, \lambda(s_1)^{n_2} \ 
 \int_0^{s_1} ds_2 \, \dot{b}_{\sigma(2)}(s_2) B(\Phi_2^{\sigma})(s_2,\lambda)\,
 k'_{\sigma(2)} \lambda(s_2)^{k'_{\sigma(2)}+n_1} \Phi_1(s_2,\lambda) \EEQ
\BEA && F_{2'}(\sigma)=4(n_2+1)(n_1+1) \int_0^t ds_1\ \lambda(s_1)^{n_2} \int_0^{s_1}
ds_2\, \dot{b}_{\sigma(2)}(s_2) \, 
B(\Phi_2^{\sigma})(s_2,\lambda) \nonumber\\
&& \qquad \qquad \qquad k'_{\sigma(2)} \lambda(s_2)^{k'_{\sigma(2)}-1} \
\dot{\lambda}(s_2) \int_0^{s_2} ds_3\, \lambda(s_3)^{n_1} \Phi_1(s_3,\lambda)
\EEA

One checks straightforwardly that $\half \partial_t F_{1'}=\frac{\partial}{\partial\lambda(t)} F_{I'}$ and $\half \partial_t F_{2'}=\frac{\partial}{\partial\lambda(t)} F_{II'}$. Hence $[\bar{\del}_1,\bar{\del}_2]$ is indeed a noise-preserving
transformation.

\bigskip

The next task is obviously to express the above Lie brackets in some appropriate basis.
The natural basis here is $(L_{n,(n_1,\ldots,n_p)}^{\dot{a},(a_1,\ldots,a_p)})$
where $n\ge -1$ (or more generally $n\in\Z$), $n_1,\ldots,n_p\ge 0 $ are positive integers,  and $a,a_1,\ldots,a_k$ are chosen in some fixed basis of time functions (for
instance among $t^l,l=0,1,\ldots$). By definition, $L_{n,(n_1,\ldots,n_p)}^{\dot{a},(a_1,\ldots,a_p)}$ acts on the trajectories $(\lambda(t))_{t\ge 0}$ as the infinitesimal
transformation $\lambda\mapsto \lambda+\eps\del\lambda$, with
\BEQ (\del\lambda)(t) :=\dot{a}(t) \lambda(t)^{n+1} \Phi^{(n_1,\ldots,n_p)}(\dot{a}_1,
\ldots,\dot{a}_p;\lambda)(t) -2(n+1)\dot{\lambda}(t) \Phi^{(n,n_1,\ldots,n_p)}(\dot{a}_0,\dot{a}_1,\ldots,\dot{a}_p;\lambda)(t)
\EEQ
(see noise-preserving condition in Definition \ref{def:noise-preserving-transformations})
where $\dot{a}_0$ is the constant function $\equiv 1$.
We refrain from computing $[L_{n,(n_1,\ldots,n_p)}^{\dot{a},(a_1,\ldots,a_p)},
L_{n',(n'_1,\ldots,n'_{p'})}^{\dot{a}',(a'_1,\ldots,a'_{p'})}]$ for general indices
$p,p'$ and correspond only elementary transformations $L_{n_i}^{\dot{a}_i}$ of the type
(\ref{eq:elementary-transformation}), corresponding to  $p,p'=0$. Computing the bracket
in the above basis yields 
\BEQ [L_{n_1}^{\dot{a_1}},L_{n_2}^{\dot{a}_2}]=(n_2-n_1)L_{n_1+n_2}^{\dot{a}_1
\dot{a}_2}-2 \left\{ (n_2+1) L_{n_1,n_2}^{\ddot{a}_1,a_2} - (n_1+1) L_{n_2,n_1}^{\ddot{a}_2,a_1} \right\}.\EEQ
The second and last terms in the above equation become very simple for $n_1,n_2=-1,0$
since $L_{n,0}^{\dot{a},b}=L_n^{\dot{a}b}$, which explains why the Schr\"odinger-Virasoro
algebra is closed under brackets. For $n_1,n_2\ge 1$, on the other hand, we get
iterated integrals of higher order and general formulas become very involved, exhibiting
sums over shuffles and permutations. Let us simply remark at this point that 
the linear span of the $L_{n,(\cdots)}^{\cdot,(\cdots)}$ with $n=-1,0,1$ is
a Lie subalgebra, just as the three-dimensional Lie algebra of finite conformal
transformations span$(L_{-1},L_0,L_1)$ is.


\section{The space-time geometry of the problem} \label{sec:space-time}


We consider in this article space-time transformations such as (\ref{eq:coordinate-change})
whose form is dictated by the {\em noise invariance condition} (\ref{eq:z=2}).
Briefly said, these are obtained by integrating time-dependent infinitesimal conformal
transformations and considering an associated transformation of the time parameter. 
As shown in the previous section, such transformations do not constitute a group for the composition of space-time
transformations, except if one restricts to indices $n=-1,0$, obtaining in this
way the Schr\"odinger-Virasoro group \cite{RogUnt}. Let us introduce the latter
smoothly in a pleasant geometric framework. 

It turns out that these transformations may be described in a coordinate-independent
setting on an arbitrary manifold $\R_+\times {\cal M}$, where $({\cal M},g)$ is any
Riemannian manifold with its metric two-form $g$. The applications we have in view
in the context of random matrices are $({\cal M},g)=(\R,d\lambda^2)$, resp.
$({\cal M},g)=(\C\cup\{\infty\},d\lambda\,  d\bar{\lambda})$, where $\lambda$ is an
eigenvalue of a Hermitian, resp. normal matrix.  Let $\Phi_t:{\cal M}\to{\cal M}$
$(t\ge 0)$ be a $C^1$ family of conformal diffeomorphisms of ${\cal M}$: by
definition, the Jacobian matrix $J(\Phi_t(m))\equiv \frac{D\Phi_t(m)}{Dm}$ is scalar.
Restricting to transformations $(\Phi_t)_{t\ge 0}$ such that $\Phi_0\equiv Id$, we
get a set $C^1_0=C^1_0(\R_+,{\mathrm{Conf}}({\cal M}))$. Consider now a world-line
$(m_t,t):=(\Phi_t(m),t)$ in ${\cal M}\times\R_+$. A non-relativistic particle living
on this world-line has {\em proper time}
\BEQ T_m(t)\equiv T(m,t)=\int_0^t ds\, |J(\Phi_s(m))|^{\alpha},  \label{eq:proper-time} \EEQ
where the dynamical scaling exponent  $\alpha$ equals $2$ if we want (\ref{eq:z=2}) to
be satisfied. This relation defines an extended space-time diffeomorphism
$\tilde{\Phi}:{\cal M}\times\R_+\to {\cal M}\times\R_+$, $\tilde{\Phi}(m,t)=(T_m(t),
\Phi_t(m))$ such that 
\BEQ \tilde{\Phi}(m,0)=(m,0),\qquad \frac{d}{dt} T(m,t)= \frac{(\Phi_t)_*(g_{ij}dx^i dx^j) (m)}{g_{ij}dx^i dx^j(m)}
=|J(\Phi_t(m))|^2.\EEQ
Alternatively, $(\Phi_t)_{t\ge 0}$ is characterized by the velocity field
$v(m,t)\equiv \frac{d\Phi_t(m,t)}{dt}$, a time-dependent vector field on $\cal M$
such that $v(\cdot,t)$ belongs for every fixed $t$ to the conformal Lie algebra ${\mathfrak{conf}}({\cal M}).$

\bigskip
\noindent Keeping to the one-dimensional case, let us now introduce the Schr\"odinger-Virasoro group in this context.
Let ${\cal M}$ be flat space $\R^d$ for some $d\ge 1$. Global conformal transformations
are simply affine transformations, i.e. compositions of rotations, scale changes
 $x\mapsto a x$ $(a\in\R)$ and translations, $x\mapsto x+v$ $(v\in\R^d)$.

\begin{itemize}
\item[(i)]  
 Let $\phi:\R_+\to\R_+$ be a $C^1$-diffeomorphism with $\phi(0)=0$. Define
 $a(t)\equiv \half \frac{\ddot{\phi}(t)}{\dot{\phi}(t)}$ and $v(x,t):=a(t)x.$
Then 
\BEQ \tilde{\Phi}(x,t)=\Big(e^{\int_0^t ds\, a(s)} x, \int_0^t ds\,  e^{2\int_0^s ds' \,   a(s')}
 \Big)=(\sqrt{\dot{\phi}}(t)x,\phi(t)).\EEQ

\item[(ii)] Let $v(x,t):=\dot{b}(t)$ for some function $b:\R_+\to\R^d$. Then
\BEQ  \tilde{\Phi}(x,t)=\left(x+\int_0^t ds\, b(s),t\right).\EEQ
\item[(iii)] Let $v(x,t):={\cal R}(t) x$ where ${\cal R}(t)\in {\mathfrak{so}}(d)$ is
an antisymmetric matrix (infinitesimal rotation). Then 
\BEQ \tilde{\Phi}(t,x)=\left(t, \overset{\longrightarrow}{\exp} \left( \int_0^t  ds\, 
{\cal R}(s)\right) x\right)\EEQ
where $\overset{\longrightarrow}{\exp}$ is the time-ordered exponential.
\end{itemize}

Composing these transformations one obtains a zero-mass representation of the so-called Schr\"odinger-Virasoro group. Considering infinitesimal transformations and restricting
to the one-dimensional case for simplicity, one gets 
\BEQ \frac{d}{d\eps} F\Big(
e^{-\half\eps \int_0^t ds\, \ddot{f}(s)} x\int_0^t ds\,  e^{-\eps\int_0^s ds' \,   \ddot{f}(s')}  \Big) \Big|_{\eps=0}=(X_{f}F)(x,t)\EEQ
and
\BEQ  \frac{d}{d\eps} F\Big(x-\eps g(t),t\Big) \Big|_{\eps=0}=(Y_g F)(x,t) \EEQ
where the vector fields
\BEQ X_f:=-f(t)\partial_t-\half \dot{f}(t) x\partial_x, \qquad 
Y_g:=-g(t)\partial_x \EEQ
make up a zero mass representation of the Schr\"odinger-Virasoro algebra, namely,
\BEQ [X_f,X_g]=X_{\dot{f}g-f\dot{g}} \EEQ
\BEQ [Y_f,X_g]=Y_{\dot{f}g-\half f\dot{g}}, \qquad [Y_f,Y_g]=0. \label{eq:sv} \EEQ
 
One recognizes elementary transformations as in (\ref{eq:elementary-transformation}),
with $X_f\equiv L_0^{\dot{f}}$ and $Y_g\equiv L_{-1}^{g}$ (see end of \S 2 for
the notation).


\section{Dynamical constraints} \label{sec:dynamical-constraints}



\subsection{Introduction and statement of main result}


If we search for a dynamical analogue of the equilibrium constraints $L_n^{eq} {\cal Z}[\tau]=0$, we must choose a dynamical functional replacing the partition function.
Clearly a substitute for the equilibrium measure is the measure $\cal Q$ on trajectories. However
(whatever its precise dependence on the parameters of the potential), 
$\cal Q$ is always normalized, viz. ${\cal Q}[1]=1$. Using
the trivial identity $\del {\cal Q}[1]=0$ does give non-trivial identities, but
${\cal Q}[1]$ is not a generating functional. Instead we consider a perturbed
evolution, 
\BEQ d\lambda_i=dB_i-\frac{\partial W[\tau]}{\partial\lambda_i} dt,\EEQ
where $W[\tau](\{\lambda_i\})\equiv W(\{\lambda_i\})+\half \sum_{k\ge 1} \tau_k \sum_i \lambda_i^k$. 
Copying the change-of-measure leading to (\ref{eq:Girsanov}), and throwing away the
second-order term in $\tau$ completing the square, $e^{-\frac{1}{4} \sum_i \left(
\partial_{\lambda_i}\left(\sum_{k\ge 0}\tau_k\lambda_i^k\right)\right)^2}$, we get a
{\em new measure}
\BEQ {\cal Q}[\tau](\{\lambda_i\})= {\cal Q}(\{\lambda_i\})   e^{-\sum_{k=1}^{+\infty} \sum_i \int_0^{+\infty} \,  \tau_k(t) k \lambda_i^{k-1}(t)
\left( d\lambda_i(t)+\left(V'(\lambda_i(t))-\sum_{j\not=i} \frac{\beta}{\lambda_i(t)-\lambda_j(t)}
\right) dt \right) }  \label{eq:Q[tau]} \EEQ

Using It\^o's formula, 
\BEQ k\lambda_i^{k-1} d\lambda_i(t)=d(\lambda_i^k)(t)-k(k-1)\lambda_i^{k-2}(t)dt,\EEQ
and the identity
\BEQ \beta\sum_{i,j\not=i} \frac{\lambda_i^{k-1}}{\lambda_i-\lambda_j}=
\frac{\beta}{2} \sum_{q=0}^{k-2}
(\pi_q\pi_{k-2-q}-\pi_{k-2}),\EEQ
we obtain  a second, more useful expression for ${\cal Q}[\tau]$, 
\BEQ {\cal Q}[\tau](\{\lambda_i\})= {\cal Q}(\{\lambda_i\}) e^{-S[\tau]},\EEQ where
\BEQ S[\tau]\equiv \sum_{k\ge 0}  \int S_k(t)\tau_k(t)\, dt, \EEQ
\BEQ \qquad
S_k(t)=\dot{\pi}_k(t)+ (\frac{\beta}{2}-1) k(k-1)\pi_{k-2}(t)+  k\sum_{l\ge 0} b_l \pi_{l+k-1}(t)
-\frac{\beta}{2} k\sum_{q=0}^{k-2} \pi_q(t) \pi_{k-2-q} (t) .\EEQ
The action $S[\tau]$ is the sum of a linear term (linearized action) 
\BEQ S^{lin}[\tau]\equiv \sum_{k\ge 0} \int S^{lin}_k(t)\tau_k(t)\, dt, \qquad 
S^{lin}_k(t)=\dot{\pi}_k(t)  + (\frac{\beta}{2}-1) k(k-1)\pi_{k-2}(t)+ k\sum_{l\ge 0} b_l \pi_{l+k-1}(t) \EEQ
and of a quadratic term, $S^{quadr}[\tau]$.

Throwing this quadratic term in turn, we finally get a linearized functional.

\begin{Definition}[generating functional] \label{def:generating-functional}
Let
\BEQ {\cal Q}^{lin}[\tau](\{\lambda_i\}):={\cal Q}(\{\lambda_i\})e^{-S^{lin}[\tau]}\EEQ
and
\BEQ {\cal Z}^{lin}[\tau]:={\cal Q}^{lin}[\tau][1]=\int d{\cal Q}^{lin}[\tau]
 (\{\lambda_i\}).
\label{eq:Z[tau]} \EEQ
\end{Definition}

For simplicity we shall from now on write $\langle \ \cdot\ \rangle_0$ instead of 
${\cal Q}(\ \cdot\ )$, viz. $\langle \ \cdot\ \rangle_\tau$ instead of 
${\cal Q}^{lin}[\tau](\ \cdot\ )$

Differentiating with respect to $\tau$ yields
\BEA && \frac{\del}{\del\tau_k(t)} e^{-S[\tau]}=-e^{-S[\tau]}\left( \dot{\pi}_k(t)
+ (\frac{\beta}{2}-1) k(k-1)\pi_{k-2}(t) + k\sum_{l\ge 0} b_l \pi_{l+k-1}(t) \right. \nonumber\\
&& \qquad\qquad\qquad\qquad \left.
-\frac{\beta}{2} k\sum_{q=0}^{k-2} \pi_q(t) \pi_{k-2-q} (t) \right),
\EEA
formally,
\BEQ \dot{\pi}_k(t) + (\frac{\beta}{2}-1) k(k-1)\pi_{k-2}(t) + k\sum_{l\ge 0} b_l \pi_{l+k-1}(t)
-k\frac{\beta}{2} \sum_{q=0}^{k-2} \pi_q(t) \pi_{k-2-q} (t)=-\frac{\del}{\del\tau_k(t)}\label{eq:non-linear-EM}\EEQ
in average (i.e. when inserted into an expectation value).
\medskip

We need to invert the linearized version of this equation,
\BEQ \dot{\pi}_k(t) + (\frac{\beta}{2}-1) k(k-1)\pi_{k-2}(t) + k\sum_{l\ge 1} b_l \pi_{l+k-1}(t)
=-\frac{\del}{\del\tau_k(t)}, \qquad k\ge 1 \label{eq:linearized-eq} \EEQ
with solution 
\BEQ \pi_k(t)=-\sum_l \int_0^t ds\, K_{kl}(t-s) \frac{\del}{\del\tau_l(s)},  \label{eq:K} \EEQ
 or, in a
mixed operator/convolutional notation, $\pi_k(t)=-(K\ast  \frac{\del}{\del\tau})_k(t)$.
Note that $K(t-s)=(K_{kl}(t-s)_{kl}$ is an upper-triangular matrix, so the sum in
(\ref{eq:K}) really ranges over $l\ge k$.
At this point we introduce the generating series 
\BEQ \widehat{\partial/\partial\tau}(z,t)\equiv \sum_{k\ge 1}
z^{-k-1} \frac{\partial}{\partial\tau_k(t)}, \  \pi(z,t)=\sum_{k\ge 0} \pi_k(t) z^{-k-1},
\EEQ
\BEQ  \tau(z,t)=\sum_{k\ge 1} k\tau_k(t) z^{k-1}, \  b(z)=\sum_{l\ge 1} b_l z^l \EEQ
Note that the zero mode $\pi_0(t)$ of the field $\pi(z,t)$ is a constant, $\pi_0(t)=N$.
In somewhat abstract terms, we use the canonical splitting of the formal series algebra
$\C[[z,z^{-1}]]$ into ${\cal A}_+\oplus{\cal A}_-\equiv \C[[z]]\oplus z^{-1} \C[[z^{-1}]]$;
each of these two subalgebras is isotropic for the scalar product 
\BEQ (u,v)=
\oint u(z)v(z)dz:= \frac{1}{2\II\pi} \int_{{\cal C}} u(z)v(z)\, dz \EEQ
given by the residue integral, where $\cal C$ is any counterclockwise simple contour
circling aroung $0$,  and ${\cal A}_+= {\cal A}_-^*$. Then
$\widehat{\partial/\partial\tau}(z),\pi(z)\in{\cal A}_-$, while $b(z)\in {\cal A}_+$.
We write quite  generally 
\BEQ u_-(z)=({\cal P}_- u)(z):=\sum_{n\le -1} u_n z^n, \qquad
u_+(z)=({\cal P}_+ u)(z):=\sum_{n\ge 0} u_n z^n \EEQ
if $u(z)=\sum_{n\in\Z} u_n z^n\in\C[[z,z^{-1}]]$ (see Appendix B). Also, we
write : $u(z)=v(z) \ \mod {\cal A}_+$ if ${\cal P}_-u={\cal P}_-v$. With these
notations, 
letting  $\dot{\pi}=\frac{\partial\pi}{\partial t}, \pi'=\frac{\partial\pi}{\partial z}$, we see that (\ref{eq:linearized-eq}) is equivalent to the following

\begin{Definition}[equation of motion] \label{def:equation-motion}
\BEQ \dot{\pi}(z)+ (\frac{\beta}{2}-1) \pi''(z)-(b\pi)'(z)=- \widehat{\partial/\partial\tau}(z) \ \ \mod {\cal A}_+.
\label{eq:4.13} \EEQ 
\end{Definition}

As mentioned in the Introduction, the original non-linearized equation of motion
(\ref{eq:non-linear-EM}) contributes to (\ref{eq:4.13}) an additive term $(\pi^2)'(z)$
which turns it into a Burgers equation, but we shall not pursue along this road.

\medskip

\noindent The solution of (\ref{eq:4.13}) is
\BEQ \pi(z,t)\equiv  -z^{-1} \int_0^t ds\, \oint dw\,  K_{t-s}(z^{-1},w) \widehat{\partial/\partial\tau}(w,s)\EEQ
or more schematically, 
\BEQ \pi(z,t)\equiv -(K\ast \widehat{\partial/\partial\tau})(z,t), \EEQ
 where (comparing with (\ref{eq:K}))  
 \BEQ K_{t-s}(z^{-1},w)=\sum_{k,l\ge 0}
z^{-k}w^l K_{kl}(t-s). \EEQ 

When $t\to 0$ we must get $K_t(z^{-1},w)\to \frac{1}{1-w/z}$, so that
$z^{-1} \oint dw\,  K_0(z^{-1},w)f(w^{-1})=f(z^{-1})$ for every $f=f(w^{-1})=a_0+a_1 w^{-1}
+a_2 w^{-2}+\ldots$.

When $\beta=2$, the equation of motion (\ref{eq:4.13}) is a transport equation which may
be solved explicitly in terms of the characteristics; as proved in
Appendix B, 
\BEQ K_t(z^{-1},w)=\frac{1}{1-w(t)/z}  \label{eq:sol-eq-motion-beta=2} \EEQ
where $w(t)\in\C[[w]]$ is the solution at time $t$ of the ordinary differential
equation $\dot{w}_t=-b(w_t)$ with initial condition $w_0\equiv w$.
When $\beta\not=2$, semi-explicit but complicated formulas for $K_t$ may be
obtained by composing the semi-group generated by $\partial_z^2: \pi\mapsto -\pi''$ with
the  semi-group generated by the transport equation $B:\pi\to (b\pi)'(z)$ through
the use of Trotter's formula, $\exp\, t\left(-(\frac{\beta}{2}-1)\partial_z^2+B \right)=\lim_{n\to \infty}
\left( \exp(-\frac{t}{n}(\frac{\beta}{2}-1)\partial_z^2) \exp(\frac{t}{n}B) \right)^n$, resulting
in a Feynman-Kac type formula which looks awful. Hence we do not write it down, but
the reader should be able to reproduce it by looking at the computations in Appendix B.

\medskip
As mentioned in the introduction, we see that n-point functions of the functions $\{\pi_k(t)\}$ may be obtained by differentiating $\cal Z$,
\BEQ \Big{\langle} \left(\int dt\,  f_1(t)\pi_{k_1}(t)\right)\cdots 
 \left(\int dt\,  f_p(t)\pi_{k_p}(t)\right) \Big{\rangle}_0 =
 \prod_{q=1}^p  \left( -\int dt\, f_q(t) (K\ast \widehat{\partial/\partial\tau})_{k_q}(t) \right)
 {\cal Z}^{lin}[\tau]  \Big|_{\tau=0}. \label{eq:n-point-functions} \EEQ

\medskip

The kernel $K$ satisfies the semi-group properties,
\BEQ \sum_{l\ge 0} K_{kl}(t-t') K_{lm}(t'-t'')=K_{lm}(t-t''), \qquad 
t>t'>t''\EEQ
or equivalently
\BEQ \oint \frac{dz'}{z'} K_{t-t'}(z^{-1},z')K_{t'-t''}((z')^{-1},z'')=
K_{t-t''}(z^{-1},z''); \label{eq:KKK} \EEQ
letting $t'\to t$ or $t'\to t''$ and differentiating we get 
the following formulas, \BEQ \frac{\partial}{\partial t} K_{km}(t)=-\sum_{l\ge 0}  kb_{l-k+1} K_{lm}(t)=
-\sum_{l\ge 0} K_{kl}(t) lb_{m-l+1}  \label{eq:semi-group-3} \EEQ
or equivalently
\BEA &&  \frac{\partial}{\partial t} K_t(z^{-1},z'')=-\oint \frac{dz'}{z'} \frac{b(z')}{z} \frac{1}{(1-z'/z)^2} K_t((z')^{-1},z'') \label{eq:Kolmogorov1}\\
&& \qquad \qquad \qquad =-\oint \frac{dz'}{z'} K_t(z^{-1},z') \frac{b(z'')}{z'} \frac{1}{(1-z''/z')^2}. \label{eq:Kolmogorov2} \EEA
In the last two equalities we used the following expression for the generator,\\
 $\sum_{k,l\ge 0} z^{-k}w^l kb_{l-k+1}=
\frac{b(w)}{z} \frac{1}{(1-w/z)^2}$. 
Following the probabilists' convention we shall refer to (\ref{eq:Kolmogorov1}),
resp. (\ref{eq:Kolmogorov2})  as the
{\em forward}, resp. {\em backward Kolmogorov equation}.

\bigskip

We now define the two bosonic fields.

\begin{Definition}[static and dynamic free bosons] \label{def:free-boson}

\begin{itemize}
\item[(i)] (static free boson)
Let, for $k\ge 1$,
\BEQ \hat{\phi}_{-k}(t):=\beta^{-1/2}k\tau_k(t),\qquad \hat{\phi}_k(t):=\beta^{1/2}
\frac{\del}{\del\tau_k(t)}\EEQ
and $\hat{\phi}_0:=0$,
\BEQ \hat{\phi}(z,t):=\sum_{k\in\Z} \hat{\phi}_k(t) z^{-k-1}.\EEQ

\item[(ii)] (dynamic free boson) 
Let, for $k\ge 1$,
\BEQ \hat{\psi}_{-k}(t):=\beta^{-1/2} k \tau_k(t),\qquad \hat{\psi}_k(t):= \beta^{1/2} (K\ast \frac{\del}{\del\tau})_k(t)\EEQ
and $\hat{\psi}_0:=-\beta^{1/2} N$,
\BEQ  \hat{\psi}(z,t)\equiv \sum_{k\in\Z} \hat{\psi}_k z^{-k-1}(t).\EEQ
\end{itemize}
\end{Definition}

Since $ (K\ast \frac{\del}{\del\tau})_k(t)$ identifies with $-\pi_k(t)$ for $k\ge 1$,
and $\pi_0(t)=\sum_i 1\equiv N$, 
the definition of the zero mode $\hat{\psi}_0$ is coherent. Then
\BEQ \hat{\psi}(z,t)\equiv \hat{\psi}_+(t,z)+\hat{\psi}_-(t,z) \EEQ
where 
\BEQ \hat{\psi}_+(z,t)=\beta^{-\half} \sum_{k\ge 1} k 
\tau_{k}(t) z^{k-1}=\beta^{-1/2}\tau(z,t) \ \in {\cal A}_+, \EEQ
\BEA \hat{\psi}_-(z,t)&=&-\beta^{1/2} \left\{ Nz^{-1}-  \sum_{k\ge 1}  (K\ast \frac{\del}{\del\tau})_k(t) z^{-k-1} \right\}=\beta^{1/2} \left\{ -Nz^{-1}+ (K\ast \widehat{\partial/\partial\tau})(z,t) \right\} \nonumber\\
&=& \beta^{1/2} \left\{ -Nz^{-1}+ z^{-1} \int_0^t ds \oint d\zeta\, K_{t-s}(z^{-1},\zeta)
\widehat{\partial/\partial\tau}(\zeta,s) \right\}.\EEA
Similarly,
\BEQ \hat{\phi}(z,t)\equiv \hat{\phi}_+(t,z)+\hat{\phi}_-(z,t) \EEQ
where 
\BEQ \hat{\phi}_+\equiv \hat{\psi}_+,\qquad \hat{\phi}_-(z,t)=\beta^{1/2}\,  \widehat{\partial/\partial\tau}(z,t).\EEQ
For further use we write down a formula regarding the time-derivative of $\hat{\psi}$,
\BEQ \partial_t(\hat{\psi}_-(z,t)) = \beta^{1/2} \left\{ \widehat{\partial/\partial\tau}(z,t)+\frac{1}{z} \int_0^t ds \oint d\zeta\, 
\partial_t(K_{t-s}(z^{-1},\zeta))\,  \widehat{\partial/\partial\tau}(\zeta,s) \right\}.
\label{eq:time-derivative-psi}\EEQ
Alternatively, from (\ref{eq:app-dtpsi-bis}),
\BEQ \partial_t(\hat{\psi}_-(z,t)) = \hat{\phi}_-(z,t)-\beta^{1/2} \int_0^t ds
\oint d\zeta\, b(\zeta) G^+_{t-s}(z^{-1},\zeta) \widehat{\partial/\partial\tau}(\zeta,s).\EEQ

\medskip

Taking commutators, we get: \medskip

\begin{Definition}[Dynamic free boson algebra] \label{def:free-boson-algebra}
 Let $G^+(t,z^{-1};t',w)=G^+_{t-t'}(z^{-1},w)$ be
the {\bf retarded propagator}, 
\BEQ G^+_{t-t'}(z^{-1},w):= {\bf 1}_{t>t'} \frac{1}{z} \frac{\partial}{\partial w} K_{t-t'}(z^{-1},w),  \label{eq:G+} \EEQ
 $G^-(t',z;t,w^{-1}):=G^+(t,w^{-1};t',z)$ the {\bf advanced propagator}, and 
\BEQ G^+_0(z^{-1},w):=\lim_{t\to 0,t>0} G^+_t(z^{-1},w)=\frac{1}{z^2(1-w/z)^2}, \EEQ
\BEQ \qquad\qquad\qquad
 G^-_0(z,w^{-1}):=\lim_{t\to 0,t<0} G^-_t(z,w^{-1})=\frac{1}{w^2(1-z/w)^2}. \EEQ
 
Then
\BEQ [\hat{\phi}(z,t),\hat{\phi}(w,t')]=\del(t-t') \left\{ G^+_0(z^{-1},w)-G^-_0(z,w^{-1})
\right\}  \EEQ
\BEQ  [\hat{\psi}(z,t),\hat{\psi}(w,t')]=G^+_{t-t'}(z^{-1},w)-G^-_{t-t'}(z,w^{-1}) \EEQ
\BEQ [\hat{\psi}(z,t),\hat{\phi}(w,t')]=G^+_{t-t'}(z^{-1},w)-\del(t-t') G^-_0(z,w^{-1}).\EEQ
\end{Definition}

\medskip
Let us give a sketchy proof. First, if $k,l\ge 0$, 
\BEQ [\hat{\psi}_{k}(t),\hat{\psi}_{-l}(t')]={\bf 1}_{t>t'} lK_{kl}(t-t'). \EEQ
Summing over Fourier components yields for $t>t'$
\BEQ  [\hat{\psi}(z,t),\hat{\phi}(w,t')]= [\hat{\psi}(z,t),\hat{\psi}(w,t')]=\sum_{k,l\ge 0} lw^{l-1} z^{-k-1} K_{kl}(t-t')=G^+_{t-t'}(z^{-1},w)\EEQ
Other commutators either vanish identically or involve a $\delta$-function.

\bigskip

{\bf Example (Hermite polynomials).} Assume $b_1=1/\sigma^2$ and $b_i=0$, $i\not=1$.  Then
the equation (\ref{eq:4.13}) reduces to $\dot{\pi}+(\frac{\beta}{2}-1)\pi''=\frac{1}{\sigma^2}(\pi+z\frac{d\pi}{dz})-\widehat{\partial/\partial\tau}(z)$. 

Consider first the case $\beta=2$.  The equation is diagonal when written in Fourier modes,
$\dot{\pi}_k=-\frac{1}{\sigma^2} k\pi_k-\frac{\partial}{\partial\tau_k}$. The solution is 
$$\pi(z,t)=\int_0^t ds\,  \sum_{k\ge 0} e^{-k(t-s)/\sigma^2} \frac{\partial}{\partial\tau(s)}
z^{-k-1}=\int_0^t ds \, \frac{1}{z} \oint dw\, \sum_{k\ge 0} (e^{-(t-s)/\sigma^2}w/z)^k 
\widehat{\partial/\partial\tau}(w,s). $$
 Hence for $t\ge 0$
\BEQ K_t(z^{-1},w)=\frac{1}{1-e^{-t/\sigma^2}w/z}, \qquad G^+_{t}(z^{-1},w)={\bf 1}_{t>0} \frac{e^{-t/\sigma^2}}{z^2(1-e^{-t/\sigma^2}w/z)^2}.  \label{eq:Hermite} \EEQ
Note that $G^+_0(z^{-1},w)=\frac{1}{z^2(1-w/z)^2}$; one retrieves the equal-time, equilibrium
OPE \\ $\phi(z,t)\phi(w,t')\sim \del(t-t')\frac{1}{(z-w)^2}$.
 
 \medskip
 
When $\beta\not=2$, an explicit expression for $K_t$, loosely related
to the Mehler kernel,  is given in Appendix B, see
(\ref{eq:Hermite-betanot=2}). 
 
\vskip 2cm

We may now state our main result. We denote by $C_c^{\infty}(\R_+^*)$ the
space of smooth functions with compact support $\subset(0,+\infty)$. (In particular,
a function in $C_c^{\infty}(\R_+^*)$ vanishes to arbitrary order at $0$).

\bigskip

\begin{Theorem}[dynamical constraints]  \label{th:main}  
Let, for  $a\in C_c^{\infty}(\R_+^*)$, 
\BEA  && L^a_{-1} :=\beta^{-1/2} \int dt \left\{  \ddot{a}(t) 
\oint \, z\hat{\psi}_t(z)\, dz  -  a(t) \oint ((\frac{\beta}{2}-1)b''+b'b)(z) \hat{\psi}_t(z)\, dz \right\}  \nonumber\\ 
&&    -\int dt \left[ \half\dot{a}(t) \oint \ :\, (\hat{\psi}(z,t))^2\, : \ dz - \half a(t) \left\{ \oint b'(z) \ :\, (\hat{\psi}(z,t))^2 \, :\  dz +  \oint 
\ :(\hat{\phi}(z,t))^2 \, : \ dz \right\}  \right]  \nonumber\\
\EEA
and

\BEA && L^a_0 := -a(t)\partial_t + \half\beta^{-1/2} \int dt \left\{  \half \dddot{a}(t)  
\oint z^2 \hat{\psi}_t(z)\, dz - 
\dot{a}(t) \oint ((\frac{\beta}{2}-1)(zb(z))''+(zb(z))'b(z)) \hat{\psi}_t(z)\, dz \right\} 
 \nonumber\\
&&   -\half \int dt\left[ \half\ddot{a}(t) \oint \ :\, (\hat{\psi}(z,t))^2 \, :\ z\, dz  
\right. \nonumber\\
&& \left. \qquad\qquad\qquad  -\half\dot{a}(t) \left\{
 \oint (zb(z))' \ :\, (\hat{\psi}(z,t))^2\, : \ dz  +  \oint z\ :\, (\hat{\phi}(z,t))^2\, :
 \ dz\right\} \right]. \nonumber\\ \label{eq:4.50} \EEA

Then
\BEQ L^a_n {\cal Z}[\tau]=0, \qquad n=-1,0.\EEQ
\end{Theorem}

The proof is elementary but somewhat lengthy.  It will take up the rest of the section.
 As it happens, see (\ref{eq:coordinate-change}), the Schr\"odinger-Virasoro transformation $Y_{\dot{a}}$,
$$\del\lambda(t)=-\dot{a}(t), \ \del t=0 $$
generates $L_{-1}^{\dot{a}}$, while the transformation $X_{a}$,
$$\del\lambda(t)=-\lambda \dot{a}(t), \ \del t=-a(t)$$
generates $L_0^{a}$. 

\medskip

\noindent The action of the time derivation $a(t)\partial_t$ is made explicit in the Appendix.


\subsection{Preliminary computations}


We show here that the variation $\del {\cal Q}^{lin}[\tau]:=\del_n {\cal Q}^{lin}[\tau]$ of the action
under the change of coordinates (\ref{eq:coordinate-change}) is the sum of four terms,
$\del_{(i)}{\cal Q}^{lin}[\tau],\ldots,\del_{(iv)}{\cal Q}^{lin}[\tau]$ which we evaluate one by one.

\medskip

 First, a straightforward extension of Girsanov's formula
yields
\BEA \del {\cal Q}^{lin}[\tau]&=&{\cal Q}^{lin}[\tau] \left( \int \del V'_i(t) \, d\lambda_i(t) +
\int \del V'_i(t) \frac{\partial W}{\partial\lambda_i}(t) \, dt \right) -
{\cal Q}^{lin}[\tau]\,  \del\left(  \sum_{k=0}^{+\infty} \tau_k(t)
S_k(t) \right) \nonumber\\
&=&  \del_{(i)}{\cal Q}^{lin}[\tau]+  \del_{(ii)}{\cal Q}^{lin}[\tau]+ \del_{(iii)}{\cal Q}^{lin}[\tau]+
 \del_{(iv)}{\cal Q}^{lin}[\tau]
\EEA
with 
\BEQ \del_{(i)}{\cal Q}^{lin}[\tau]={\cal Q}^{lin}[\tau] \int \del V'_i(t) \, d\lambda_i(t), \qquad
\del_{(ii)}{\cal Q}^{lin}[\tau]={\cal Q}^{lin}[\tau] \int \del V'_i(t) \frac{\partial W}{\partial\lambda_i}(t)\, dt, \EEQ
\BEQ \del_{(iii)}{\cal Q}^{lin}[\tau]=- {\cal Q}^{lin}[\tau] \int \del V'_i(t) \sum_{j\not=i}
\frac{\beta}{\lambda_i(t)-\lambda_j(t)}\, dt, \qquad
\del_{(iv)}{\cal Q}^{lin}[\tau]=-{\cal Q}^{lin}[\tau]\,  \del\left(  \sum_{k=0}^{+\infty} \tau_k(t)
S_k(t) \right).\EEQ
We consider separately each of these four terms.

\begin{itemize}
\item[(i)] The quantity $\del V'_i$ is given by the total variation formula (\ref{eq:3.5}) as a sum of four terms. Though the third and the fourth one vanish for $n=-1,0$, we
evaluate them to some point and shall use those computations in another article. By It\^o's formula,
\BEQ \lambda_i^{n+1}(t) d\lambda_i(t)=\frac{1}{n+2} d(\lambda_i^{n+2})(t)-(n+1)\lambda_i^n
(t) dt \EEQ
hence (by integration by parts)
\BEA &&   \sum_i \int \ddot{a}(t) \lambda_i^{n+1}(t) d\lambda_i(t)  = - \frac{1}{n+2}
\int \dddot{a}(t) \pi_{n+2}(t)\,  dt - (n+1) \int \ddot{a}(t) \pi_n (t)\,  dt
\nonumber\\
&&= -\frac{1}{n+2} \dddot{a}(t)  \oint \pi(z) z^{n+2}\, dz+ \ddot{a}(t) \oint \pi'(z) z^{n+1}\,
dz.
\EEA

 The second and fourth terms are similar,
\BEA && \sum_i \int \dot{a} \left[ \sum_{l\ge 0} b_l (n+l+1) \lambda_i^{n+l}  - (\frac{\beta}{2}-1)(n+1)n\lambda_i^{n-1} \right] d\lambda_i \nonumber\\
&& =-\int \ddot{a} \left[ \sum_{l\ge 0} b_l \pi_{n+l+1} -(\frac{\beta}{2}-1) (n+1)\pi_n \right] dt+ \nonumber\\
&& \ \ -  \int \dot{a} \left[ \sum_{l\ge 0} b_l (n+l+1)(n+l)\pi_{n+l-1}
-(\frac{\beta}{2}-1) (n+1)n(n-1)\pi_{n-2} \right] dt\nonumber\\
&& =-\ddot{a} \left( \oint b(z)\pi(z) z^{n+1}\, dz + (\frac{\beta}{2}-1) \oint z^{n+1} \pi'(z)\, dz \right) \nonumber\\
&& \qquad \qquad - \dot{a} \left( \oint b(z) \pi''(z) z^{n+1}\, dz + (\frac{\beta}{2}-1) \oint z^{n+1} \pi'''(z)\, dz\right) 
  \nonumber\\ \label{eq:4.30} \EEA

For the third term, we remark similarly that
\BEQ d\left( \sum_{q=0}^{n-1} \pi_q \pi_{n-q} \right)=2\sum_{q=0}^{n-1} 
\sum_i d\lambda_i 
(q+1)\lambda_i^q \pi_{n-q-1} + \sum_{q=0}^{n-1} \sum_i \frac{\partial^2}{\partial\lambda_i^2}
(\pi_{q}\pi_{n-q}) dt.\EEQ
We compute
\BEA   \sum_{q=0}^{n-1} \sum_i \frac{\partial^2}{\partial\lambda_i^2}
(\pi_{q}\pi_{n-q}) &=&  2\left( \sum_{q=1}^{n-1} q(q-1) \pi_{q-2}\pi_{n-q} +
 q(n-q) \pi_{q-1}\pi_{n-q-1} \right) \nonumber\\
 &=& 2(n+1)  \sum_{q=1}^{n-1} (q-1) \pi_{q-2}\pi_{n-q} \nonumber\\ \EEA
Hence
\BEA && \beta\sum_{q=0}^{n-1}\sum_i \int \dot{a}(t)
(q+1)\lambda_i^q \pi_{n-1-q}\   d\lambda_i(t)= - \frac{\beta}{2} \int \ddot{a}(t) 
\sum_{q=0}^{n-1}  \pi_q(t) \pi_{n-q}(t) \, dt -
\nonumber\\ && \qquad 
-\beta (n+1) \int \dot{a}(t) \sum_{q=1}^{n-1} (q-1) \pi_{q-2}(t)
\pi_{n-q}(t) \, dt \nonumber\\
&& \qquad \qquad = -\frac{\beta}{2} \ddot{a} \oint \pi^2(z)\,  z^{n+1} \, dz -
\beta \dot{a} \oint (\pi \pi')'(z) \, z^{n+1}\, dz.
\EEA

\item[(ii)] 
We  now evaluate $\int dt\, \del V'_i V'(\lambda_i)$ . It is a linear combination of
terms of the type (with $\phi(t)=\dot{a}(t)$ or $\ddot{a}(t)$)
$b_l\int dt\, \phi \sum_i \lambda_i^{m+l}=b_l\int dt\, \phi \pi_{m+l}$ and
$b_l \int dt\, \phi \sum_i \lambda_i^{k+l} \pi_{q-k} =b_l\int dt\, \phi \pi_{q+l}$.
Summing up all four terms, we get a $\beta$-independent contribution,
\BEA &&  \ddot{a} \sum_{l\ge 0} b_l \pi_{l+n+1}+\dot{a} \sum_{l,l'\ge 0} b_l b_{l'} (l'+n+1) 
\pi_{l+l'+n} +\beta \dot{a}\sum_{l\ge 0} b_l \sum_{q=0}^{n-1}  (q+1) \pi_{n+l-1}
\nonumber\\
&& \qquad \qquad 
-\dot{a} \sum_{l\ge 0} b_l \ \cdot\ (\frac{\beta}{2}-1) (n+1)n \pi_{n+l-1}  \nonumber\\
&& \qquad =\ddot{a} \sum_{l\ge 0} b_l \pi_{l+n+1}+\dot{a} \left[ \sum_{l,l'\ge 0} b_l b_{l'} (l'+n+1) 
\pi_{l+l'+n}  + \frac{\beta}{2} (n+1)n \sum_{l\ge 0} b_l   \pi_{n+l-1} \right] \nonumber\\
&& =\ddot{a} \oint b(z)\pi(z) z^{n+1}\, dz+ \dot{a} \left(- \oint b(z) (b\pi)'(z)z^{n+1}\,
 dz + \frac{\beta}{2} \oint (b\pi)''(z) z^{n+1}\, dz\right). \nonumber\\ \label{eq:4.34}
 \EEA

Comparing (\ref{eq:4.34}) with (\ref{eq:4.30}), we see that the first terms sum up to zero.

\item[(iii)] We now evaluate $-\sum_{i,i'\not=i} \int dt\,
\del V'_i \frac{\beta}{\lambda_i-\lambda_{i'}}$. First
\BEA  -\frac{\beta}{2} \sum_{i,i'\not=i} \int dt\, \phi \frac{\lambda_i^m-\lambda_{i'}^m}{\lambda_i-\lambda_{i'}} &=& - \frac{\beta}{2}
\int dt\, \phi(t) \sum_{q=0}^{m-1} \left(\pi_q(t)
\pi_{m-1-q}(t)-\pi_{m-1}(t)\right) \nonumber\\
&=& \frac{\beta}{2} \int dt \, \phi(t) \left( m\pi_{m-1}(t)-\sum_{q=0}^{m-1} \pi_q(t) \pi_{m-1-q}(t) \right) ;  \nonumber\\ \EEA

summing up the contributions of the terms 1,2,4 in (\ref{eq:3.5}), we get

\BEA &&  \frac{\beta}{2}\ddot{a}\left( (n+1)\pi_n- \sum_{q=0}^{n} \pi_q
\pi_{n-q}\right) + \frac{\beta}{2}\dot{a} \sum_{l\ge 0} b_l (l+n+1) \left( (l+n)\pi_{l+n-1} -\sum_{q=0}^{l+n-1} \pi_q
\pi_{l+n-1-q} \right) - \nonumber\\
&&   -\frac{\beta}{2}\dot{a} \ \cdot\  (\frac{\beta}{2}-1) (n+1)n \left( (n-1)\pi_{n-2} -\sum_{q=0}^{n-2}
\pi_q \pi_{n-2-q} \right)  \nonumber\\
&&= -\frac{\beta}{2}\ddot{a} \oint (\pi'(z)+\pi^2(z)) z^{n+1}\, dz+ \frac{\beta}{2}\dot{a} \left(\oint b(z) (\pi''(z)+(\pi^2(z))') z^{n+1}\, dz \right.\nonumber\\ && \qquad  \left.
+ \frac{\beta}{2}(\frac{\beta}{2}-1)  \oint (\pi'''(z)+(\pi^2(z))'') z^{n+1}\, dz \right).  \nonumber\\
\EEA 

The third term contributes
\BEA && -\frac{\beta^2}{2} \int dt\, \dot{a} \sum_{q=0}^{n-1} (q+1) \sum_{i,j,i'\not=i} \frac{(\lambda_i^q-\lambda_{i'}^q) \lambda_j^{n-1-q}}{\lambda_i-\lambda_{i'}} \nonumber\\
&& =-\frac{\beta^2}{2}  \int dt\, \dot{a}
\sum_{q=1}^{n-1}  (q+1) \sum_{p=0}^{q-1}   \sum_{i,j,i'\not=i}\lambda_i^p \lambda_{i'}^{q-1-p}
\lambda_j^{n-1-q} \nonumber\\
&&= -\frac{\beta^2}{2}  \int dt\, \dot{a} \Big\{ 2 \sum_{q+r+s=n-2} (q+1) \pi_q \pi_r\pi_s - 
\sum_{k=0}^{n-2} (k+2)(k+1) \pi_{k} \pi_{n-2-k} \Big\} \nonumber\\
&&= -\frac{\beta^2}{2} \int dt\, \dot{a} \left\{ -2\oint \pi'(z) \pi^2(z) z^{n+1}\, dz  - \oint \pi''(z) \pi(z) z^{n+1}\, dz \right\},
 \EEA
  
including a term of order 3 (which does not appear for $n\le 1$ however). 

\item[(iv)] 
Finally, we must compute the variation $-\del_n\left( \int_0^{+\infty} dt\, \sum_{k=0}^{+\infty} \tau_k(t)
S_k(t) \right)=-\del S^{lin}$ under the change of coordinates (\ref{eq:coordinate-change}) for $n=-1,0.$  First we have  
\BEQ \del_n(\pi_k(t))=-\dot{a}(t) k\pi_{k+n}(t), \qquad \del_n(\dot{\pi}_k)=
\frac{d}{dt}(\del_n(\pi_k))=-\dot{a} k\dot{\pi}_{k+n} -\ddot{a} k\pi_{k+n}.\EEQ
Recall we have defined $\tau(z)\equiv \sum_{k\ge 1} k\tau_k z^{k-1}.$
 Thus, for $n=-1$
 
\BEA &&  -\del_{-1} S^{lin}=\dot{a} \left\{ \sum_{k\ge 0} k\tau_k \sum_{l\ge 0} b_l (l+k-1)
\pi_{l+k-2} + \sum_{k\ge 0} k\tau_k \dot{\pi}_{k-1}  + (\frac{\beta}{2}-1)
\sum_{k\ge 0} k(k-1)(k-2)\tau_k \pi_{k-3} \right\} \nonumber\\ && \qquad \qquad
\qquad\qquad\qquad\qquad  + \ddot{a}
\sum_{k\ge 0} k\tau_k \pi_{k-1}\nonumber\\
&&= \dot{a} \left\{ -\oint \tau(z)b(z)\pi'(z) \, dz + \oint \tau(z)\dot{\pi}(z)
+ (\frac{\beta}{2}-1) \oint \tau(z)\pi''(z)\, dz \, dz
\right\} + \ddot{a} \oint \tau(z)\pi(z)\, dz. \label{eq:del-1Slin} \EEA

When $n=0$ we must add to a term similar to (\ref{eq:del-1Slin}) a   contribution
$\del_0^{time}(S)$ due to the time change; letting $\tilde{t}=t-\eps\del t=t+2\eps a$ be the new time coordinate, we get 
\BEA  && -\eps\del_0^{time}(\int dt\, \tau_k(t) S_k^{lin}(t)) = -\eps \del_0^{time}\left( \int dt\, \tau_k(t)
\left(\dot{\pi}_k(t)+ (\frac{\beta}{2}-1)k(k-1)\pi_{k-2}(t) \right)
\right.\nonumber\\ && \qquad \left.  + k\sum_{l\ge 0} b_l \int dt\, \tau_k(t)\pi_{l+k-1}(t) \right) = \int dt \, \tau_k(t) S_k^{lin}(t)  - \int d\tilde{t} (\tau_k(\tilde{t})-2\eps a
\dot{\tau}_k(\tilde{t})) \frac{d\pi_k}{d\tilde{t}}(\tilde{t})  \nonumber\\
&& \qquad - 
\int d\tilde{t}\, (1-2\eps\dot{a}(\tilde{t}))(\tau_k(\tilde{t})-2\eps a \dot{\tau}_k(\tilde{t})) \left\{ k\sum_{l\le 0}b_l \pi_{l+k-1}(\tilde{t}) +
(\frac{\beta}{2}-1)k(k-1)\pi_{k-2}(\tilde{t}) \right\}  \nonumber\\
&&\qquad = 2\eps \left( \int dt\, a(t) \dot{\tau}_k(t) \dot{\pi}_k(t) + 
\int dt\, (\dot{a}(t)\tau_k(t)+a(t)\dot{\tau}_k(t))\ \cdot
\right.\nonumber\\ && \qquad\qquad \qquad  \ \cdot\ \left. \left\{ k\sum_{l\ge 0} b_l\pi_{l+k-1}(t)
+ (\frac{\beta}{2}-1)k(k-1)\pi_{k-2}(t) \right\} \right). \nonumber\\
\EEA

Hence

\BEA && -\del_0 S^{lin}=\dot{a} \left\{ \sum_{k\ge 0} k\tau_k \sum_{l\ge 0} b_l (l+k-1)
\pi_{l+k-1} + \sum_{k\ge 0} k\tau_k \dot{\pi}_{k} + (\frac{\beta}{2}-1) \sum_{k\ge 0} k(k-1)(k-2)\tau_k \pi_{k-2} 
 \right\} \nonumber\\ && \qquad  + \ddot{a} \sum_{k\ge 0} k\tau_k \pi_k +   2\dot{a} \left( \sum_{k\ge 0} k\tau_k \sum_{l\ge 0} b_l \pi_{l+k-1} +  (\frac{\beta}{2}-1)
 \sum_{k\ge 0} k(k-1)\tau_k\pi_{k-2} \right)  \nonumber\\ && \qquad 
 + 2a \left\{ \sum_{k\ge 0} \dot{\tau}_k (\dot{\pi}_k+ (\frac{\beta}{2}-1)k(k-1)\pi_{k-2})  + \sum_{k\ge 0} k\dot{\tau}_k
 \sum_{l\ge 0} b_l\pi_{l+k-1} \right\} \nonumber\\ 
&&= \ddot{a} \oint \tau(z)\pi(z) z\, dz+ \dot{a}  \left\{ -\oint \tau(z) b(z) (z\pi(z))' \, dz+ \oint \tau(z)(\dot{\pi}(z)+ (\frac{\beta}{2}-1) \pi''(z)) z \, dz  \right.\nonumber\\ && \qquad \qquad \qquad \left.+ 2 \oint \tau(z)b(z)\pi(z)\, dz\right\}
  -2a\partial_t \label{eq:del0Slin}
\EEA
where $a\partial_t=a(t)\partial_t$ is the time derivation acting on the coefficients
 of the functional ${\cal Z}[\tau]$ (see Appendix B). The term $\oint \tau(z)\pi''(z)z\, dz$ in 
 (\ref{eq:del0Slin}) is equal to the sum  $\sum_{k\ge 0} k(k-1)(k-2)\tau_k\pi_{k-2}+
2 \sum_{k\ge 0}k(k-1)\tau_k \pi_{k-2}$. 

\end{itemize}


\subsection{Schr\"odinger-Virasoro transformations}


We may  finally collect all contributions to obtain generators of transformations, denoted
by 
$L_{-1}^{\dot{a}}$ and $L_0^{2a}$. Recall $\pi(z,t)\equiv -\beta^{-1/2} \hat{\psi}_-(z,t)$.

\begin{itemize}
\item[(i)] $(n=-1)$ 
Collecting all terms in $\del_{(i)}$, $\del_{(ii)}$, $\del_{(iii)}$ and $\del_{(iv)}$, we
get a term $L_{-1,lin}$ linear in $(\tau,\partial/\partial\tau)$, plus a term $L_{-1,quadr}$ which is
quadratic,
\BEQ L_{-1}^{\dot{a}} \equiv \int dt\, \left\{ L^{\dot{a}}_{-1,lin}(t)+L^{\dot{a}}_{-1,quadr}(t)\right\}, \EEQ
\BEA L_{-1,lin}^{\dot{a}}(t) &=&-\dddot{a} \pi_1 + (\frac{\beta}{2}-1) \dot{a} \oint b(z)\pi''(z)\,  dz - \dot{a}
\oint b(z) (b\pi)'(z)\, dz \nonumber\\
&=& \beta^{-1/2} \left\{ \dddot{a} \oint z\hat{\psi}(z)\, dz - \dot{a} \oint 
((\frac{\beta}{2}-1) b''(z)+b(z)b'(z)) \hat{\psi}(z)\, dz \right\}  \nonumber\\
 \EEA

\BEA &&  L_{-1,quadr}^{\dot{a}}(t)= \dot{a} \left\{ \frac{\beta}{2} \oint b(z)(\pi^2)'(z)\, dz  -
\oint \tau(z) b(z) \pi'(z)\, dz + \oint \tau(z) \dot{\pi}(z) \, dz \right.\nonumber\\
&& \qquad \qquad \qquad  \left.  +
(\frac{\beta}{2}-1) \oint \tau(z) \pi''(z)\, dz \right\} +
\ddot{a} \oint \tau(z)\pi(z)\, dz. \EEA

Using (\ref{eq:4.13}) and taking into
account the fact that the subalgebras ${\cal A}_+$, ${\cal A}_-$ are isotropic, we get
\BEA && L_{-1,quadr}^{\dot{a}}(t)=\dot{a} \left\{ -\frac{\beta}{2} \oint b'(z) (\pi^2)(z)\, dz - \oint \tau(z)b(z)\pi'(z)\, dz + (\frac{\beta}{2}-1) \oint \tau(z) \pi''(z)\, dz \right. \nonumber\\
&& \qquad \qquad \qquad  \left. + \oint \tau(z) ((b\pi)'(z)-\widehat{\partial/\partial\tau}(z)-
(\frac{\beta}{2}-1)\pi''(z))\, dz \right\}  +
\ddot{a} \oint \tau(z)\pi(z)\, dz \nonumber\\
&&= \dot{a} \left\{ -\oint b'(z) ( \frac{\beta}{2}(\pi^2)(z)-\tau(z)\pi(z))\, dz  -\oint \tau(z)\widehat{\partial/\partial\tau}(z) \, dz \right\} 
  + \ddot{a} \oint \tau(z)\pi(z)\, dz \nonumber\\
&&=-\dot{a} \left\{ \half \oint b'(z) \  (\beta^{1/2}\pi(z)-\beta^{-1/2} \tau(z))^2
 \ \, dz  \right. \nonumber\\
&& \qquad \qquad \left. + \half \oint \, :\, (\beta^{1/2}\widehat{\partial/\partial\tau}(z) + \beta^{-1/2} \tau(z))^2\, :\  \, dz  \right\}  \nonumber\\
&& \qquad\qquad\qquad\qquad  -\half \ddot{a} \oint\,   (\beta^{1/2}\pi(z)-\beta^{-1/2} \tau(z))^2  \, dz\nonumber\\
&&=-\dot{a} \left( \half \oint b'(z) \ :\, (\hat{\psi}(z))^2 \, :\  dz +
  \half \oint \ :(\hat{\phi}(z))^2 \, : \ dz \right) 
-\half \ddot{a} \oint \ :\, (\hat{\psi}(z))^2\, : \ dz \nonumber\\
\EEA

\item[(ii)] $(n=0)$ One finds
\BEQ L_0^{2a}\equiv-2a\partial_t + \int dt \left\{ L_{0,lin}^{2a}(t)
+L_{0,quadr}^{2a}(t) \right\}, \EEQ
\BEA L_{0,lin}^{2a}(t) &=& (\frac{\beta}{2}-1)\ddot{a}\pi_0-\half \dddot{a}  \pi_2+
\dot{a} \left\{  (\frac{\beta}{2}-1)\oint b(z)\pi''(z)z\, dz - \oint b(z) (b\pi)' z\, dz
\right\} \nonumber\\
&=& (\frac{\beta}{2}-1)N\ddot{a}+\beta^{-1/2} \left\{ \half \dddot{a} \oint z^2
\hat{\psi}(z) dz - \dot{a} \oint ((\frac{\beta}{2}-1) (zb(z))''+ (zb(z))'b(z))\hat{\psi}_t(z)\, dz \right\} \nonumber\\
\EEA
(note that the first term, a total derivative, disappears after integration in
(\ref{eq:4.50})); 

\BEA && L_{0,quadr}^{2a}(t)=\ddot{a} \oint z(\tau\pi-\frac{\beta}{2}\pi^2)(z) \, dz+ \dot{a} \left\{ \frac{\beta}{2} \oint zb(z)(\pi^2)' (z)\, dz + (\frac{\beta}{2}-1)
\oint \tau(z)\pi''(z)z\, dz + \right.\nonumber\\ && \left. \ 
\oint z\tau(z) ((b \pi)'(z)-\widehat{\partial/\partial\tau}(z)-(\frac{\beta}{2}-1)
\pi''(z))  \, dz    - \oint z \tau(z) b(z) \pi'(z) \, dz + \oint \tau(z)b(z)\pi(z)\, dz
 \right\}  \nonumber\\
 &&= \ddot{a} \oint z(\tau\pi-\frac{\beta}{2}\pi^2)(z) \, dz+ \dot{a} \left\{ \oint (zb(z))' (\tau\pi-\frac{\beta}{2}\pi^2)(z)\, dz -\oint z\tau(z) \widehat{\partial/\partial\tau}(z) \, dz \right\}
 \nonumber\\
 &&= -\half \ddot{a} \oint \ :\, (\hat{\psi}(z))^2 \, :\ z\, dz -\half \dot{a} \left\{
 \oint (zb(z))' \ :\, (\hat{\psi}(z))^2\, : \ dz+ \oint z\ :\, (\hat{\phi}(z))^2\, :
 \ dz\right\}. \nonumber\\
 \EEA
\end{itemize}


\subsection{Commutators: the quadratic part}


We prove in this paragraph that $(L_{-1,quadr}^a,L_{0,quadr}^a)_{a\in C_c^{\infty}(\R_+^*)}$ provide a
zero mass representation of the Schr\"odinger-Virasoro algebra, see (\ref{eq:sv}):

\begin{Theorem}
\BEQ \Big[\int L_{0,quadr}^f(t)\, dt, \int L_{0,quadr}^g(t')\, dt'\Big]=
\int L_{0,quadr}^{\dot{f}g-f\dot{g}}(t)\, dt, \EEQ
\BEQ \Big[\int L_{-1,quadr}^f(t)\, dt, \int L_{0,quadr}^g(t')\, dt'\Big]=
\int L_{-1,quadr}^{\dot{f}g-\half f\dot{g}}(t)\, dt,
\ \  \Big[\int L_{-1,quadr}^f(t)\, dt, \int L_{-1,quadr}^g(t')\, dt' \Big]=0.\EEQ
\end{Theorem}

In order to keep computations to a reasonable length, we consider commutators of
the functionals
\BEA && A^{quadr}_v(f):=-2v'(z) f(t)\partial_t -\half\int dt\, \ddot{f}(t) \oint \ :\, (\hat{\psi}_t(z))^2 \, :\ v(z)\, dz  \nonumber\\ && \qquad\qquad  -\half\int dt\, \dot{f}(t) \left\{
 \oint (v(z)b(z))' \ :\, (\hat{\psi}_t(z))^2\, : \ dz  +  \oint v(z)\ :\, (\hat{\phi}_t(z))^2\, :
 \ dz\right\} \nonumber\\ \EEA
with $v(z)=1$ or $z$. Note that $A_v^{quadr}(f)=L_{-1,quadr}^{\dot{f}}$ for $v(z)=1$, and
$A_v^{quadr}(f)=L_{0,quadr}^{2f}$ for $v(z)=z$. The {\em non-differential part} of
$A_v^{quadr}(f)$, $\bar{A}_v^{quadr}(f):=A_v^{quadr}(f)+2v'(z) f(t)\partial_t$, is
by definition $A_v^{quadr}(f)$ shorn of its differential part $-2v'(z)f(t)\partial_t$.  The first computations are valid
for an arbitrary function $v\in\C[[z]]$, but at some point we must restrict to
$v(z)=1$ or $z$, which are the only cases needed. We want to prove:
\BEQ [A_z^{quadr}(f),A_z^{quadr}(g)]=4\int  L_{0,quadr}^{\dot{f}g-f\dot{g}}(t)\, dt, \EEQ
\BEQ [A_1^{quadr}(f),A_z^{quadr}(g)]= 2 \int L_{-1,quadr}^{\ddot{f}g-\half \dot{f}\dot{g}}(t)\, dt,
\qquad [A_1^{quadr}(f),A_1^{quadr}(g)]=0.\EEQ

\medskip

 Using the commutation relations of the
boson algebra, we find for $u,v\in \C[[z]]$ 
\BEA && \left[ \half \oint u(z) \, :\, (\hat{\psi}_t(z))^2\, :\,  \, dz, \half \oint   v(w) \, :\,(\hat{\phi}_{t'}(w))^2\, : \, dw\right] \nonumber\\
&& \qquad = \oint \oint dz dw\, u(z)v(w) \, :\, \hat{\psi}_t(z)\hat{\phi}_{t'}(w)\, :\,  G^+_{t-t'}(z^{-1},w)
\nonumber\\ && \qquad \qquad
- \del(t-t') \oint \oint dz dw\, u(z)v(w) \, :\, \hat{\psi}_t(z) \hat{\phi}_t(w)\, :\,  G_0^-(z,w^{-1}); \EEA
\BEA &&  \left[ \half \oint u(z)\, :\, (\hat{\psi}_t(z))^2\, : \, dz, \half \oint  v(w)  \, :\, (\hat{\psi}_{t'}(w))^2\, : \, dw\right] \nonumber\\ 
&& \qquad = \oint \oint dz dw\, u(z)v(w)
\, :\, \hat{\psi}_t(z)\hat{\psi}_{t'}(w)\, : \ 
  (G^+_{t-t'}(z^{-1},w)-G^-_{t-t'}(z,w^{-1}))  \nonumber\\ \EEA

\BEA && \left[ \half \oint u(z) \, :\, (\hat{\phi}_t(z))^2 \, :\, dz, 
\half \oint v(w) \, :\, (\hat{\phi}_{t'}(w))^2 \, :\, dw \right] \nonumber\\
&& \qquad \qquad  =
\oint dz \oint dw \, u(z) v(w) \, :\, \hat{\phi}_t(z)\hat{\phi}_{t'}(w)\, :\, 
(G_0^+(z^{-1},w)-G_0^-(z,w^{-1})) \nonumber\\ \EEA

hence
\BEQ [A_u^{quadr}(f),A_v^{quadr}(g)]=\sum_{i=1}^8 C_i(f,g),\EEQ
where:

\begin{itemize}
\item[(i)]  (contribution of the commutator $[\psi,\psi]$)

\BEA C_1(f,g) &=& \oint dz dw\, \int dt \int_0^t dt' \, ( \dot{f}(t) (ub)'(z)+
u(z)\ddot{f}(t)) v(w)\ddot{g}(t') \, :\, \hat{\psi}_t(z)\hat{\psi}_{t'}(w)\, :\,  G^+_{t-t'}(z^{-1},w) \nonumber\\
&-&  \oint dz dw\, \int dt' \int_0^{t'} dt\,  u(z)\ddot{f}(t)   ( \dot{g}(t') (vb)'(w)+
v(w)\ddot{g}(t'))  \, :\, \hat{\psi}_t(z)\hat{\psi}_{t'}(w)\, :\,  G^-_{t-t'}(z,w^{-1}) \nonumber\\
&\equiv & C_{1,1}(f,g)+C_{1,2}(f,g)+C_{1,3}(f,g)\EEA
where (by integrating by parts with respect to $t'$ or $t$, and using the fundamental relations (\ref{eq:P},\ref{eq:P-cor1},\ref{eq:P-cor2}))

\BEA && (1)\qquad C_{1,1}(f,g)= \nonumber\\
&&  \qquad = - \oint dz \, \int dt  \, ( \dot{f}(t) (ub)'(z)+
u(z)\ddot{f}(t)) \dot{g}(t) \hat{\psi}_t(z)  ({\cal P}_-(v\hat{\psi}_t))'(z)  \nonumber\\
&& \qquad \qquad + \oint  dw\, \int dt'   \dot{f}(t') (\dot{g}(t')(vb)'(w)+v(w)\ddot{g}(t')) \hat{\psi}_{t'}(w)
 ({\cal P}_-(u\hat{\psi}_{t'}))'(w)  \nonumber\\
\EEA

If $u=v$ then the two terms in $\dot{f}(t)\dot{g}(t)$ cancel each other, and there
remains only
\BEQ C_{1,1}(f,g)=- \oint dz \int dt\,  (\ddot{f}(t)\dot{g}(t)-\dot{f}(t)\ddot{g}(t))
({\cal P}_+(u\hat{\psi}_t))(z) ({\cal P}_- (u\hat{\psi}_t))'(z).
\label{eq:C11u=v} \EEQ
Otherwise we may assume that $u(z)=1$, $v(w)=w$, from which $({\cal P}_-(v\hat{\psi}_t))'(z)= (v {\cal P}_-(\hat{\psi}_t))'(z)=z({\cal P}_-\hat{\psi}_t)'(z)+({\cal P}_t
\hat{\psi}_t)(z)$ (observe that the first equality is wrong if $v(w)=w^{n+1}$, $n\ge 1$)
and 
\BEA && C_{1,1}(f,g)= \nonumber\\
&& \ - \oint dz\, z \int dt\,  (\ddot{f}(t)\dot{g}(t)-\dot{f}(t)\ddot{g}(t))
({\cal P}_+(\hat{\psi}_t))(z) ({\cal P}_- (\hat{\psi}_t))'(z) \label{eq:C11u=1v=z1} \\
&& - \int dt\, \dot{f}(t)\dot{g}(t) \left\{ \oint dz \, b'(z) \hat{\psi}_t(z) ({\cal P}_-\hat{\psi}_t)(z) - \oint dz \, b(z) \hat{\psi}_t(z) ({\cal P}_-\hat{\psi}_t)'(z) \right\}
\nonumber \\
&& -\frac{1}{2}\int dt\, \ddot{f}(t)\dot{g}(t) \oint dz\,   \, :\, (\hat{\psi}_t(z))^2\, : 
\label{eq:C11u=1v=z3} \EEA
In the last line we have used: $$\oint dz\, :\, (\hat{\psi}_t(z))^2\, : \ =2\oint dz\, ({\cal P}_+\hat{\psi}_t)(z) ({\cal P}_-\hat{\psi}_t)(z)=2\oint dz\, \hat{\psi}_t(z) ({\cal P}_-\hat{\psi}_t)(z).$$

\BEA && (2)\qquad C_{1,2}(f,g)= \nonumber\\
&& - \beta^{1/2} \oint dzdw\, \int dt\int_0^t dt'  ( \dot{f}(t) (ub)'(z)+ u(z)
\ddot{f}(t)) v(w)\dot{g}(t') \hat{\psi}_t(z) G^+_{t-t'}(z^{-1},w) \widehat{\partial/\partial\tau}(w,t')-sym. \nonumber\\
&&= -\oint dzdw\, \int dt\int_0^t dt'  ( \dot{f}(t) (ub)'(z)+ u(z)
\ddot{f}(t)) v(w)\dot{g}(t') \hat{\psi}_t(z) G^+_{t-t'}(z^{-1},w) \hat{\phi}_{t'}(w) - sym. ,  \nonumber\\
\EEA
 a contribution due to the first term in the right-hand side of  (\ref{eq:time-derivative-psi}); "-sym". indicates, here as in the following computations,  a similar term with the kernel $G^-_{t-t'}$ in factor;

\BEA &&  (3) \qquad C_{1,3}(f,g)= -\beta^{1/2}   \int dt\int_0^t dt'\int_0^{t'}ds    \oint dz\,  (\dot{f}(t)(ub)'(z)+u(z)\ddot{f}(t))
\dot{g}(t') \hat{\psi}_t(z) \nonumber\\
&& \qquad \qquad \frac{1}{z} \partial_{t'} \left( \oint \frac{dw}{w}\,  v(w) \partial_w(K_{t-t'}(z^{-1},w))
K_{t'-s}(w^{-1},\zeta) \right) \widehat{\partial/\partial\tau}(\zeta,s) - sym.  \nonumber\\
&&= -\beta^{1/2}  \int dt\int_0^t dt'\int_0^{t'}ds    \oint dz\,  (\dot{f}(t)(ub)'(z)+u(z)\ddot{f}(t))
\dot{g}(t') \hat{\psi}_t(z) \nonumber\\
&& \qquad \qquad \frac{1}{z} \partial_{t'} \oint \frac{dw}{w}\,  (v(w)b'(w)-v'(w)b(w)) \partial_w(K_{t-t'}(z^{-1},w)) K_{t'-s}(w^{-1},\zeta)\widehat{\partial/\partial\tau}(\zeta,s) - sym. \nonumber\\
&& =- \int dt\int_0^t dt' \oint dzdw\,  (\dot{f}(t)(ub)'(z)+u(z)\ddot{f}(t))
\dot{g}(t') \nonumber\\ && \qquad\qquad    (v(w)b'(w)-v'(w)b(w)) \, :\, \hat{\psi}_t(z)\hat{\psi}_{t'}(w)\, :\,  G^+_{t-t'}(z^{-1},w) - sym. \nonumber\\
\EEA
by (\ref{eq:technical-lemma});

\item[(ii)] (contribution of the commutator $[\psi,\psi]$, continued)
\BEA && C_2(f,g)= \int dt\int_0^t  dt' \oint dzdw\,  (\dot{f}(t)(ub)'(z)+u(z)\ddot{f}(t))
\dot{g}(t') (vb)'(w)\,  :\, \hat{\psi}_t(z)\hat{\psi}_{t'}(w)\,:\,  G^+_{t-t'}(z^{-1},w) \nonumber\\
&& \qquad - \int dt'\int_0^{t'} dt \oint dzdw\,  \dot{f}(t)(ub)'(z)(\dot{g}(t')
(vb)'(w)+v(w)\ddot{g}(t'))
\,  :\, \hat{\psi}_t(z)\hat{\psi}_{t'}(w)\,:\,  G^-_{t-t'}(z,w^{-1}) \nonumber\\
\EEA

\item[(iii)] (non-$\del$ contribution of the commutator $[\psi,\phi]$ for $t\not=t'$)
\BEA && C_3(f,g)= \int dt\int_0^t  dt' \oint dzdw\,  (\dot{f}(t)(ub)'(z)+u(z)\ddot{f}(t))
v(w) \dot{g}(t') \, :\, \hat{\psi}_t(z)\hat{\phi}_{t'}(w)\, :\,  G^+_{t-t'}(z^{-1},w) \nonumber\\
&& \qquad -  \int dt'\int_0^{t'} dt \oint dzdw\, \dot{f}(t)u(z)(\dot{g}(t')
(vb)'(w)+v(w)\ddot{g}(t))\, :\,  \hat{\phi}_t(z) \psi_{t'}(w)\, :\,   G^-_{t-t'}(z,w^{-1})
\nonumber\\  \EEA

\item[(iv)] ($\delta$-contribution of the commutator $[\psi,\phi]$)
\BEA C_4(f,g) &=&- \int dt \oint dzdw\,  (\dot{f}(t)(ub)'(z)+u(z)\ddot{f}(t))
v(w)\dot{g}(t)\hat{\phi}_{t}(w)  \hat{\psi}_t(z) G_0^-(z,w^{-1}) \nonumber\\
&+&  \int dt \oint dzdw\,  u(z)\dot{f}(t) (\dot{g}(t) (vb)'(w)+v(w)\ddot{g}(t))
 \hat{\phi}_t(z) \hat{\psi}_{t}(w) G_0^+(z^{-1},w) \nonumber\\
&=& - \int dt \oint dz\,  (\dot{f}(t)(ub)'(z)+u(z)\ddot{f}(t))
\dot{g}(t)  ({\cal P}_+(v\hat{\phi}_t))'(z) \hat{\psi}_t(z) \nonumber\\
&+&  \int dt \oint dw\,  \dot{f}(t)( \dot{g}(t) (vb)'(w)+v(w)\ddot{g}(t))
 ({\cal P}_+(u\hat{\phi}_t))'(w)  \hat{\psi}_t(w) \nonumber\\ \EEA

This term is parallel to the term $C_{1,1}(f,g)$, to the analysis of which we
refer. In the following expressions, we use the fact that ${\cal P}_+ \hat{\phi}_t=
{\cal P}_+\hat{\psi}_t$. When $u=v$ we find
\BEQ C_4(f,g)=- \oint dz \int dt\,  (\ddot{f}(t)\dot{g}(t)-\dot{f}(t)\ddot{g}(t))
 ({\cal P}_+ (u\hat{\psi}_t))'(z) ({\cal P}_-(u\hat{\psi}_t))(z).
\label{eq:C4u=v} \EEQ
Otherwise we may assume that $u(z)=1$, $v(w)=w$, from which 
\BEA && C_{4}(f,g)= \nonumber\\
&& \ - \oint dz\, z \int dt\,  (\ddot{f}(t)\dot{g}(t)-\dot{f}(t)\ddot{g}(t))
 ({\cal P}_+ (\hat{\psi}_t))'(z) ({\cal P}_-(\hat{\psi}_t))(z) \label{eq:C4u=1v=z1} \\
&& - \int dt\, \dot{f}(t)\dot{g}(t) \left\{ \oint dz \, b'(z)  ({\cal P}_+\hat{\psi}_t)(z) \hat{\psi}_t(z) - \oint dz \, b(z)  ({\cal P}_+\hat{\psi}_t)'(z) \hat{\psi}_t(z) \right\}
\nonumber \\
&& -\frac{1}{2}\int dt\, \ddot{f}(t)\dot{g}(t) \oint dz\,   \, :\, (\hat{\psi}_t(z))^2\, : 
\label{eq:C4u=1v=z3} \EEA

\item[(v)] ($\del$-contribution due to the commutator $[\phi,\phi]$)

This term  clearly vanishes when $u=v$. Hence we may assume that $u(z)=1$, $v(w)=w$,
in which case
\BEA C_5(f,g)&=&\int dt\, \dot{f}(t)\dot{g}(t) \oint dz\, \oint dw\, w \ :\, 
\hat{\phi}_t(z)\hat{\phi}_t(w)\, :\ (G_0^+(z^{-1},w)-G_0^-(z,w^{-1})) \nonumber\\
&=& \int dt\, \dot{f}(t)\dot{g}(t) \oint dw\, w \ :\, \hat{\phi}_t(w) \left\{ 
({\cal P}_+\hat{\phi}_t)'(w)+ ({\cal P}_-\hat{\phi}_t)'(w) \right\}\, : \nonumber\\
&=& -\half \int dt\, (\dot{f}\dot{g})(t) \oint dw\, :\, (\hat{\phi}_t(w))^2\, : \EEA

\item[(vi)] asssume $v(w)=w$: by the results of Appendix B, in particular, (\ref{eq:app-fdtgtau}, \ref{eq:app-fdtgammaddtau}, \ref{eq:f(t)dtpsi-},\ref{eq:f(t)dtpsi-2})
\BEA && C_6(f,g) =2 \Big[ \bar{A}_u^{quadr}(f), -g(t)\partial_t \Big] \nonumber\\
&&=-2
 \int dt\, \oint dz\, ( \dot{f}(t) (ub)'(z)+u(z) \ddot{f}(t)) \nonumber\\
 && \qquad \qquad \qquad :\,  \hat{\psi}_t(z) \
\cdot\  \left\{ (g(t)\partial_t\cdot \hat{\psi}_-(z,t))+(g(t)\partial_t\cdot \hat{\psi}_+(z,t)) \right\}\,:  \nonumber\\
&& -2 \int dt\, \oint dz\, u(z) \dot{f}(t)\ :\,  \hat{\phi}_t(z) \ 
\left\{ (g(t)\partial_t \cdot\hat{\phi}_-(z,t))+ (g(t)\partial_t \cdot\hat{\phi}_+(z,t))
\right\}\, : \nonumber\\
&&=-2\left(  \int dt\, \oint dz\, ( \dot{f}(t) (ub)'(z)+u(z) \ddot{f}(t)) (\hat{\psi}_+(z,t)+
\hat{\psi}_-(z,t)) g(t)(\partial_t \hat{\psi}_-)(z,t) \right. \nonumber\\
&& \left. \qquad +\int dt\, \oint dz\, ( \dot{f}(t) (ub)'(z)+u(z) \ddot{f}(t)) \hat{\psi}_t(z)
 \int_0^t dt' \, \dot{g}(t') \oint dw\, b(w) G^+_{t-t'}(z^{-1},w) \hat{\psi}_{t'}(w)
 \right)
  \nonumber\\
&&-2 \left( -\int dt\, \oint dz\, g(t) (\dot{f}(t) (ub)'(z)+u(z) \ddot{f}(t))
\hat{\psi}_+(z,t) (\partial_t\hat{\psi}_-)(z,t)  \right. \nonumber\\
&& \left. \qquad \qquad -\int dt \, \oint dz\, \left\{ \frac{d}{dt}(g\dot{f})(t)
(ub)'(z) + u(z) \frac{d}{dt}(g\ddot{f})(t) \right\}
\hat{\psi}_+(z,t)  \hat{\psi}_-(z,t) \right) \nonumber\\
&& -2 \left( -\int dt \, \oint dz\, g(t) u(z) \dot{f}(t) (\partial_t \hat{\phi})(z,t)
\hat{\phi}_-(z,t) \right. \nonumber\\
&& \left. \qquad \qquad - \int dt\, \oint dz\, g(t) u(z) \ddot{f}(t) \hat{\phi}_t(z)
\hat{\phi}_-(z,t) \right) \nonumber\\
&& -2 \left( -\int dt\, \oint dz\, g(t) u(z) \dot{f}(t)  \hat{\phi}_+(z,t)  (\partial_t \hat{\phi})(z,t)
\right. \nonumber\\
&& \left. \qquad \qquad -\int dt\, \oint dz\, u(z) \frac{d}{dt} (g\dot{f})(t) 
 \hat{\phi}_+(z,t)  \hat{\phi}_t(z)\right) \nonumber\\
&& =: C_6^{nonloc}(f,g)+C_6^{loc}(f,g)
     \EEA
 where    
 
\BEA &&  C_6^{nonloc}(f,g):=-2\int dt\, \oint dz\, ( \dot{f}(t) (ub)'(z)+u(z) \ddot{f}(t)) \hat{\psi}_t(z) \nonumber\\
&& \qquad \qquad 
 \int_0^t dt' \, \dot{g}(t') \oint dw\, b(w) G^+_{t-t'}(z^{-1},w) \hat{\psi}_{t'}(w)
 \nonumber\\ \EEA
is a non-local term, and
  
\BEA && C_6^{loc}(f,g)=\int dt \oint dz\, \left\{ \frac{d}{dt} (g\dot{f})(t) (ub)'(z)
+u(z) \frac{d}{dt}(g\ddot{f})(t) \right\} \, : \, (\hat{\psi}_t(z))^2 \, : \nonumber\\
&& \qquad -\int dt\oint dz\, u(z) \frac{d}{dt}(g\dot{f})(t) \, :\, (\hat{\phi}_t(z))^2\, :
\nonumber\\
&& + 2 \int dt\oint dz\, g(t) \ddot{f}(t) u(z) \, : \ (\hat{\phi}_t(z))^2\, : \nonumber\\
&& + 2\int dt\oint dz\, u(z) \dot{g}(t)\dot{f}(t) \hat{\phi}_+(z,t) \hat{\phi}_-(z,t)  
\nonumber\\
&& =\int dt \oint dz\, \left\{ \frac{d}{dt} (g\dot{f})(t) (ub)'(z)
+u(z) \frac{d}{dt}(g\ddot{f})(t) \right\} \, : \, (\hat{\psi}_t(z))^2 \, : \nonumber\\
&& \qquad + \int dt\oint dz\, u(z) (g\ddot{f})(t) \, :\, (\hat{\phi}_t(z))^2\, :
\EEA

\item[(vii)] if $u(z)=z$,  $C_7(f,g):=2[-f(t)\partial_t,\bar{A}_v^{quadr}(g)]$
is "sym." of (vi);

\item[(viii)] finally, assuming $u(z)=z$ and $v(w)=w$, 
\BEQ C_8(f,g)=4[-f(t)\partial_t,-g(t)\partial_t]=4(f(t)\dot{g}(t)-\dot{f}(t)g(t))\partial_t.\EEQ

\bigskip
\end{itemize}

Let us now sum up the different contributions. We leave out the differential
term (viii) which is as expected. Note that $C_{1,2}+C_3\equiv 0$ in all
cases. Also, 
\BEA && C_{1,3}(f,g)+C_2(f,g)=2 \int dt\int_0^t dt' \oint dz dw\, 
(\dot{f}(t) (ub)'(z) + u(z) \ddot{f}(t)) \dot{g}(t') (v'b)(w) \nonumber\\
&& \qquad  \, :\, 
\hat{\psi}_t(z) \hat{\psi}_{t'}(w)\, :\ G^+_{t-t'}(z^{-1},w) - sym. \nonumber\\ 
\EEA
is exactly compensated by $C_6^{nonloc}(f,g)+C_7^{nonloc}(f,g)$. 

Now all remaining terms $(C_{1,1},C_4,C_5,C_6^{loc}, C_7^{loc})$ are local functionals
of the fields $\hat{\psi},\hat{\phi}$, i.e. are expressed as some integral
$\int dt\oint dz F(z,t,\hat{\psi}(z,t),\hat{\phi}(z,t))$.

\medskip

Assume first $u=v$. Then  $C_5=0$, $C_{1,1}+C_4=0$ (see (\ref{eq:C11u=v}),(\ref{eq:C4u=v})). Thus
 $[A_1^{quadr}(f),A_1^{quadr}(g)]=0$,
\BEA &&  [A_z^{quadr}(f),A_z^{quadr}(g)]=C_6^{loc}(f,g)+C_7^{loc}(f,g) \nonumber\\
&& \qquad = \int dt\oint dz\, \left\{ \frac{d}{dt}(g\dot{f}-f\dot{g})(t) (zb(z))' +
z \frac{d}{dt}(g\ddot{f}-f\ddot{g})(t) \right\} \ :\, (\hat{\psi}_t(z))^2\, :
\nonumber\\
&& \qquad \qquad + \int dt \oint dz \, z (g\ddot{f}-f\ddot{g})(t)\ :\, (\hat{\phi}_t(z))^2
\, : \nonumber\\
&&= 4 \int dt\, L_{0,quadr}^{\dot{f}g-f\dot{g}}(t). \EEA

\bigskip
Assume now $u(z)=1,v(w)=w$. Then  (\ref{eq:C11u=1v=z1},\ref{eq:C4u=1v=z1}) sum up to
\BEA &&  - \int dt\,  (\ddot{f}(t)\dot{g}(t)-\dot{f}(t)\ddot{g}(t))
\oint dz\, z\ :\,  \left\{ \hat{\psi}_t(z) (({\cal P}_-\hat{\psi}_t)'(z) +({\cal P}_+\hat{\psi}_t)'(z))  \right\} \, : \nonumber\\
&& \qquad \qquad   = \half \int dt\,  (\ddot{f}(t)\dot{g}(t)-\dot{f}(t)\ddot{g}(t)) \oint  dz\, \, :\, (\hat{\psi}_t(z))^2 \, : \EEA
Adding this to (\ref{eq:C11u=1v=z3},\ref{eq:C4u=1v=z3}) yields
\BEA && -\half \int dt \, \frac{d}{dt}(\dot{f}\dot{g})(t)
\, :\, (\hat{\psi}_t(z))^2\, : \nonumber\\
&& \qquad\qquad -\int dt \dot{f}(t)\dot{g}(t) \left\{ \oint dz\, b'(z) \, :(\hat{\psi}_t(z))^2\, : - \oint dz\, b(z)\ \, : \hat{\psi}_t(z) (\hat{\psi}_t)'(z)
\, :   \right\} 
\nonumber\\
&&=-\half \int dt \, \frac{d}{dt}(\dot{f}\dot{g})(t)
\, :\, (\hat{\psi}_t(z))^2\, : -\frac{3}{2} \int dt \dot{f}(t)\dot{g}(t)  \oint dz\, b'(z) \ 
:\, (\hat{\psi}_t(z))^2\, :
\EEA

To the latter expression we must still add
\BEQ C_5(f,g)=-\half \int dt \oint dz (\dot{f}\dot{g})(t) \, : \, (\hat{\phi}_t(z))^2\, : \EEQ
and
\BEA && C_6^{loc}(f,g)=\int dt\oint dz \, \left\{ \frac{d}{dt}(g\dot{f})(t) b'(z)+
\frac{d}{dt}(g\ddot{f})(t) \right\}\ :\, (\hat{\psi}_t(z))^2\, :
\nonumber\\
&& \qquad \qquad +\int dt \oint dz \, (g\ddot{f})(t)\ :\, (\hat{\phi}_t(z))^2\, : 
\nonumber\\ \EEA

Adding all terms yields as expected
\BEA && 2L_{-1,quadr}^{\ddot{f}g-\half \dot{f}\dot{g}} = - \int dt\oint dz
\, \left\{ (\ddot{f}g-\half \dot{f}\dot{g})(t) b'(z) + \frac{d}{dt} (\ddot{f}g
-\half \dot{f}\dot{g}) \right\} \ :\, (\hat{\psi}_t(z))^2\, : \nonumber\\
&& \qquad \qquad -\int dt\oint dz\, (\ddot{f}g-\half \dot{f}\dot{g})(t)\ :\, 
(\hat{\phi}_t(z))^2\ : \nonumber\\ \EEA


\subsection{Commutators: the linear contribution}


Copying what we did in the last subsection, we set
\BEQ A^{lin}_u(f):=\beta^{-1/2} \int dt \left\{ \dddot{f}(t) \oint  (\int u)(z) \hat{\psi}_t(z)\, dz
- \dot{f}(t) \oint ((\frac{\beta}{2}-1)(ub)''(z) + (ub)'(z)b(z)) \hat{\psi}_t(z)\, dz \right\} \EEQ
for $u(z)=1,z$, with $(\int u)(z)=z$, resp. $\frac{z^2}{2}$ when $u(z)=1$, resp. $z$.
In coherence with the quadratic parts, $A_u^{lin}(f)=L^{\dot{f}}_{-1,lin}$ for
$u(z)=1$, and $A_u^{lin}(f)=L_{0,lin}^{2f}$ for $u(z)=z$. Since obviously
$[A^{lin}_u(f),A^{lin}_v(g)]=0$, we must prove:
\BEQ [A_z^{lin}(f),A_z^{quadr}(g)]-(f\leftrightarrow g)=4 \int L^{\dot{f}g-f\dot{g}}_{0,lin}(t)\, dt, \EEQ
\BEQ [A_1^{lin}(f), A_z^{quadr}(g)]-[A_z^{lin}(f),A_1^{quadr}(g)]=2\int L^{\ddot{f}g-
\half \dot{f}\dot{g}}_{-1,lin}(t)\, dt, \qquad [A_1^{lin}(f),A_1^{quadr}(g)]-(f\leftrightarrow g)=0.\EEQ

For $u(z)=1,z$, $v(w)=1,w$, we find in general:
\BEQ [A_u^{lin}(f),A_v^{quadr}(g)]=\sum_{i=1}^4 D_i(f,g), \EEQ
 with:

\begin{itemize}
\item[(i)]
\BEA &&  D_1(f,g)=\left[A_u^{lin}(f), -\half \int dt' \oint dw\, v(w) \ddot{g}(t')
\, :\, (\hat{\psi}(w,t'))^2\, : \right] \nonumber\\
&& = -\beta^{-1/2} \int dt \int_0^t dt' \oint dz \, dw\  \left( (\int u)(z)\dddot{f}(t)-((\frac{\beta}{2}-1)(ub)''(z)+(ub)'(z) b(z))  \dot{f}(t) \right)  \nonumber\\
&& \qquad \qquad \qquad \ddot{g}(t')v(w) \hat{\psi}_{t'}(w) G^+_{t-t'}(z^{-1},w) 
\nonumber\\
&& = -\int dt\int_0^t dt'\int_0^{t'} ds \oint dz \, dw\, d\zeta\,  \left(z (\int u)(z)\dddot{f}(t)-((\frac{\beta}{2}-1)(ub)''(z)+(ub)'(z) b(z)) \dot{f}(t) \right) \nonumber\\
&& \qquad\qquad\qquad \ddot{g}(t')v(w) G^+_{t-t'}(z^{-1},w) \frac{1}{w} K_{t'-s}(w^{-1},\zeta) \widehat{\partial/\partial\tau}(\zeta,s)
\nonumber\\
&& = D_{1,1}(f,g)+D_{1,2}(f,g)+D_{1,3}(f,g), \EEA
where (by integrating by parts)

\BEA &&  (1)\qquad   D_{1,1}(f,g)=-\beta^{-1/2}\int dt \oint dz \, dw   \left( (\int u)(z)\dddot{f}(t)-((\frac{\beta}{2}-1)(ub)''(z)+(ub)'(z) b(z))\dot{f}(t)\right)
\nonumber\\
&&\qquad\qquad\qquad\qquad     \dot{g}(t)v(w)
   G^+_0(z^{-1},w)\hat{\psi}_t(w) \nonumber\\
&& \qquad  = \beta^{-1/2} \int dt\oint dz\,   \left( (\int u)(z)\dddot{f}(t)-((\frac{\beta}{2}-1)(ub)''(z)u+(ub)'(z) b(z))\dot{f}(t)\right) \nonumber\\
&& \qquad \qquad \qquad \dot{g}(t) \left( {\cal P}_-(v\hat{\psi}_t)\right)'(z) 
\EEA
   
\BEA && (2)\qquad      D_{1,2}(f,g)=\int dt\int_0^t ds \oint dz \,  dw\,  d\zeta \nonumber\\
&& \qquad \qquad  \left( (\int u)(z)\dddot{f}(t)-((\frac{\beta}{2}-1)(ub)''(z)+(ub)'(z) b(z)) \dot{f}(t) \right)
\nonumber\\
&& \qquad \qquad  \qquad  \dot{g}(s)v(w)
  G^+_{t-s}(z^{-1},w) \frac{1}{w}K_0(w^{-1},\zeta)\widehat{\partial/\partial\tau}(\zeta,s)  \nonumber\\
&& \qquad =    \int dt\int_0^t ds \oint dz \, \oint d\zeta \ \left( (\int u)(z)\dddot{f}(t)-((\frac{\beta}{2}-1)(ub)''(z)+(ub)'(z) b(z)) \dot{f}(t) \right) \nonumber\\
&& \qquad \qquad\qquad\qquad \dot{g}(s) v(\zeta)
   G^+_{t-s}(z^{-1},\zeta) \widehat{\partial/\partial\tau}(\zeta,s) ;
   \EEA
   
\BEA && (3) \qquad    D_{1,3}(f,g)=\int dt\int_0^t ds \int_s^t dt' 
\oint dz\,  dw\,  d\zeta \nonumber\\
&& \qquad \qquad \left( (\int u)(z)\dddot{f}(t)-((\frac{\beta}{2}-1)(ub)''(z)+(ub)'(z) b(z)) \dot{f}(t)\right) \nonumber\\
&& \qquad \qquad\qquad \frac{1}{z} \dot{g}(t') \partial_{t'} \left(\oint \frac{dw}{w}\, v(w) \partial_w(K_{t-t'}(z^{-1},w))
K_{t'-s}(w^{-1},\zeta) \right) \widehat{\partial/\partial\tau}(\zeta,s)  \nonumber\\
&&= 
\int dt  \int_0^t dt'\oint dz \, d\zeta  \ \left( (\int u)(z)\dddot{f}(t)-((\frac{\beta}{2}-1)(ub)''(z)+(ub)'(z) b(z)) \dot{f}(t)\right) \nonumber\\ 
&&  \qquad \qquad \dot{g}(t')  \oint \frac{dw}{w}  (v(w)b'(w)-v'(w)b(w)) G^+_{t-t'}(z^{-1},w) \hat{\psi}_{t'}(w) \nonumber\\ \EEA
using (\ref{eq:technical-lemma});

\item[(ii)]
\BEA &&  D_2(f,g)=\left[A_u^{lin}(f), -\half \int dt' \oint dw\,  \dot{g}(t')(vb)'(w)
\, :\, (\hat{\psi}(w,t'))^2\, : \right] \nonumber\\ 
&&= -\int dt\int_0^t dt' \oint dz \, dw\, d\zeta\nonumber\\
&& \qquad \qquad   \left( (\int u)(z)\dddot{f}(t)-((\frac{\beta}{2}-1)(ub)''(z)+(ub)'(z) b(z)) \dot{f}(t)
\right) (vb)'(w) \nonumber\\
&& \qquad \qquad\qquad  \dot{g}(t') G^+_{t-t'}(z^{-1},w) \frac{1}{w} \hat{\psi}_{t'}(w);
\nonumber\\
\EEA

\item[(iii)]
\BEA &&  D_3(f,g)=\left[A_u^{lin}(f), -\half \int dt' \oint dw\,  \dot{g}(t')v(w)
\, :\, (\hat{\phi}(w,t'))^2\, : \right] \nonumber\\ 
&&=-\int dt \int_0^t dt'  \oint dz dw\nonumber\\
&& \qquad\qquad \left( (\int u)(z)\dddot{f}(t)-((\frac{\beta}{2}-1)(ub)''(z)+(ub)'(z) b(z)) \dot{f}(t) \right)
\nonumber\\
&& \qquad\qquad\qquad \dot{g}(t') v(w) G^+_{t-t'}(z^{-1},w) \widehat{\partial/\partial\tau}(w,t'); \EEA

\item[(iv)] 
\BEQ  D_4(f,g)=[A_u^{lin}(f), -2g(t)\partial_t]=: D_{4,1}^{loc}(f,g)+D_{4,2}(f,g)+
D_4^{nonloc}(f,g), \EEQ where:

\BEA && D_{4,1}^{loc}(f,g)=2\beta^{-1/2}
\int dt \, \dddot{f} (t)  \oint (\int u)(z) g(t) (\partial_t\hat{\psi}_-)(z,t)\, dz 
\nonumber\\
&&=-2\beta^{-1/2}\int dt \, \frac{d}{dt}(g\dddot{f})(t)  \oint (\int u)(z) \hat{\psi}_-(z,t)\, dz; 
\EEA
\BEA &&  D_{4,2}^{loc}(f,g)=-2\beta^{-1/2}\int dt\,   \dot{f}(t) \oint ((\frac{\beta}{2}-1)(ub)''(z)+(ub)'(z)b(z)) g(t) (\partial_t \hat{\psi}_-)(z,t) \, dz \nonumber\\
&& = 2 \beta^{-1/2}\int dt\,  \frac{d}{dt}(g \dot{f})(t) \oint ((\frac{\beta}{2}-1)(ub)''(z)+(ub)'(z)b(z)) \hat{\psi}_-(z,t) \, dz;
\EEA

\BEA && D_4^{nonloc}(f,g)=
2\int dt \int_0^t dt' \oint dz \int dw \left\{ \dddot{f}(t) (\int u)(z) - \dot{f}(t) ( (\frac{\beta}{2}-1)(ub)''(z)+(ub)'(z)b(z)) \right\}  \nonumber\\
&& \qquad\qquad\qquad \dot{g}(t') b(w) G^+_{t-t'}(z^{-1},w) \hat{\psi}_{t'}(w). \EEA

\end{itemize}

\medskip

Let us now add the different contributions. 
First, $D_{1,2}+D_3\equiv 0$, $D_{1,3}+D_2+D_4^{nonloc}\equiv 0$, leaving out
only  local contributions,  $D_{1,1}(f,g), D_4^{loc}(f,g)$ and their symmetric
counterparts. 

Assume first $u(z)=v(z)=1$ or $z$. Then
\BEA && D_{1,1}(f,g)-(f\leftrightarrow g)=\beta^{-1/2} \int dt\, (\dddot{f}(t) \dot{g}(t)
-\dot{f}(t)\dddot{g}(t)) \oint dz\ (\int u)(z) ({\cal P}_-(u\hat{\psi}_t))'(z)\nonumber\\
&& \qquad \qquad = -\beta^{-1/2} \int dt\, (\dddot{f}(t) \dot{g}(t)
-\dot{f}(t)\dddot{g}(t)) \oint dz\, u^2(z) \hat{\psi}_t(z).\EEA
In particular, if $u(z)=v(z)=1$, this is equal to
\BEQ -N \int dt \, (\dddot{f}(t) \dot{g}(t)
-\dot{f}(t)\dddot{g}(t))=-N \int dt\, \frac{d}{dt} (\ddot{f}(t)\dot{g}(t)-\dot{f}(t)
\ddot{g}(t)) =0.\EEQ
Since there is not $D_4$-term in that case, we have proved:
 $[A_1^{lin}(f),A_1^{quadr}(g)]-(f\leftrightarrow g)=0$. The reader may easily check
 that one also gets the correct formula for $[A_z^{lin}(f),A_z^{quadr}(g)]-(f\leftrightarrow g)$.
 
There remains the case  $u(z)=1$, $v(w)=w$. Then
\BEA && D_{1,1}(f,g)-sym. = \beta^{-1/2} \int dt \left\{ 
(\dddot{f}\dot{g})(t) \oint z ({\cal P}_-(z\hat{\psi}_t))'(z)\, dz -
(\dot{f}\dddot{g})(t) \oint \frac{z^2}{2} ({\cal P}_- \hat{\psi}_t)'(z)\, dz \right\}
\nonumber\\
&& - \beta^{-1/2} \int dt\, (\dot{f}\dot{g})(t) \left\{ (\frac{\beta}{2}-1) b''(z)+b'(z)
b(z)) ({\cal P}_-(z\hat{\psi}_t))'(z) -  \right. \nonumber\\
&& \qquad \qquad \qquad \left. -((\frac{\beta}{2}-1) (zb(z))'' +
(zb(z))'b(z)) ({\cal P}_-\hat{\psi}_t)'(z) \right\} \nonumber\\
&&=-\beta^{-1/2} \int dt\, (\dddot{f}\dot{g}-\dot{f}\dddot{g})(t) \oint z\hat{\psi}_t(z)\, dz \nonumber\\
&& \qquad \qquad -3\beta^{-1/2} \int dt \, (\dot{f}\dot{g})(t) ((\frac{\beta}{2}-1)
b''(z)+b'(z)b(z)) \hat{\psi}_t(z)\, dz,\EEA
from which the reader may easily check the remaining bracket, 
$[A_1^{lin}(f), A_z^{quadr}(g)]$ \\ $-[A_z^{lin}(f),A_1^{quadr}(g)]$.


\subsection{A detailed example: the Hermite case}


We compute once again commutators for the sake of the reader in a simple case (Hermite
polynomials, $\beta=2$) using
Fourier modes.
Assume as in Example 1 that $b_1=1/\sigma^2$ and $b_i=0$, $i\not=1$ and let 
$\beta=2$. Then
\BEQ  L^a_{-1} =\int\, dt  \left\{ (\frac{\dot{a}}{\sigma^4}-\dddot{a})(t)\pi_1(t)  - \half(
\frac{\dot{a}}{\sigma^2} + \ddot{a})(t) \oint \ :\, (\hat{\psi}(z,t))^2\, : \ dz -\half \dot{a}(t)  \oint 
\ :(\hat{\phi}(z,t))^2 \, : \ dz \right\}  
\EEQ
and

\BEA && L^a_0 = \int \, dt \left\{ ( \frac{\dot{a}}{\sigma^4} -\half \dddot{a} )(t) \pi_2(t)+ \frac{4N}{\sigma^2}
\dot{a} 
- 2a(t)\partial_t -\half(2\frac{\dot{a}}{\sigma^2}+\ddot{a})(t) \oint \ :\, (\hat{\psi}(z,t))^2 \, :\ z\, dz  \right.\nonumber\\
&& \left. \qquad \qquad\qquad -\half \dot{a}(t)  \oint z\ :\, (\hat{\phi}(z,t))^2\, :
 \ dz \right\} \EEA

with $\half \oint \hat{\phi}^2(z,t)=\sum_{k\ge 2} k\tau_k(t) \frac{\del}{\del
\tau_{k-1}(t)}$ and
\BEQ \half \oint \hat{\psi}^2(z,t) \, dz = -N\tau_1(t)+\sum_{k\ge 2} k\tau_k(t)
\int_0^t ds\, e^{-(k-1)(t-s)/\sigma^2} \frac{\del}{\del\tau_{k-1}(s)} \EEQ

We first compute Lie brackets and prove that $(L_{-1}^a,L_0^a)_{a\in C^{\infty}}$
{\em provide a zero mass representation of the Schr\"odinger-Virasoro algebra.}
 Let $L_{-1,lin}(a),L_{-1,quadr}(a)$, resp. $L_{0,lin}(a),L_{0,quadr}(a)$ be the linear and quadratic parts of $L_{-1}^a$, resp. $L_0^a$ as in the previous paragraph.  Using the relations
in the dynamic boson algebra, we find
\BEA && \left[\oint \hat{\psi}^2(z,t)\, dz,\oint \hat{\phi}^2(w,t')\,dw\right]=
{\bf 1}_{t>t'} \sum_{k\ge 2} k(k-1) \tau_k(t) e^{-(k-1)(t-t')/\sigma^2} \frac{\del}{\del \tau_{k-2}(t')} \nonumber\\ && \qquad \qquad  + \del(t-t') \left\{ 2N\tau_2(t) - \sum_{k\ge 2} k(k-1) \tau_k(t)
\int_0^t ds\,  e^{-(k-2)(t-s)/\sigma^2} \frac{\del}{\del\tau_{k-2}(s)} \right\}; \nonumber\\ \EEA
for $t>t'$, 
\BEA &&    \left[\oint \hat{\psi}^2(z,t)\, dz,\oint \hat{\psi}^2(w,t')\,dw\right]= 
-2N\tau_2(t) e^{-(t-t')/\sigma^2} + \nonumber\\
&& \qquad \qquad
\sum_{k\ge 2} k(k-1)\tau_k(t) e^{-(k-1)(t-t')/\sigma^2} \int_0^{t'} ds\,  e^{-(k-2)(t'-s)/\sigma^2}
\frac{\del}{\del \tau_{k-2}(s)}.\EEA

From this we get
\BEQ [L_{-1,quadr}(f),L_{-1,quadr}(g)] \equiv \sum_{i=1}^6 (C_i(f,g)-C_i(g,f)), \EEQ
with (following the same scheme as in the previous subsection):
\begin{itemize}
\item[(i)] (contribution of the commutator $[\psi,\psi]$)
  \BEA C_1(f,g) &=&\int dt\ (\frac{\dot{f}(t)}{\sigma^2}+ \ddot{f}(t))
 \sum_k k(k-1)\tau_k(t) e^{-(k-1)t/\sigma^2} \nonumber\\ && \qquad\qquad\qquad \int_0^t dt'\, e^{(k-1)t'/\sigma^2} \ddot{g}(t') \left( \int_0^{t'} ds\,  e^{-(k-2)(t'-s)/\sigma^2} \frac{\del}{\del\tau_{k-2}(s)}
 \right) \nonumber\\
 &\equiv& C_{1,1}(f,g)+C_{1,2}(f,g)+C_{1,3}(f,g)  \EEA
where (by integration by parts)
\BEQ C_{1,1}(f,g)= \int dt\ (\frac{\dot{f}(t)}{\sigma^2}+ \ddot{f}(t))
 \sum_k k(k-1)\tau_k(t) \dot{g}(t) \int_0^t ds\, e^{-(k-2)(t-s)/\sigma^2}
 \frac{\del}{\del\tau_{k-2}(s)}; \EEQ
\BEQ C_{1,2}(f,g)=-\int dt\ (\frac{\dot{f}(t)}{\sigma^2}+ \ddot{f}(t))
 \sum_k k(k-1)\tau_k(t)  \int_0^t dt'\, e^{-(k-1)(t-t')/\sigma^2} \dot{g}(t')  \frac{\del}{\del\tau_{k-2}(t')}; \EEQ
\BEA &&  C_{1,3}(f,g)=-\int dt\ (\frac{\dot{f}(t)}{\sigma^2}+ \ddot{f}(t))
 \sum_k k(k-1)\tau_k(t)  \nonumber\\ && \qquad \qquad \int_0^t dt'\, e^{-(k-1)(t-t')/\sigma^2}\frac{\dot{g}(t')}{\sigma^2}  \left( \int_0^{t'} ds\, e^{-(k-2)(t'-s)/\sigma^2}  \frac{\del}{\del\tau_{k-2}(s)} \right) \nonumber\\
 \EEA

\item[(ii)] (contribution of the commutator $[\psi,\psi]$, continued)
\BEA && C_2(f,g) =\int dt\ (\frac{\dot{f}(t)}{\sigma^2}+ \ddot{f}(t))
 \sum_k k(k-1)\tau_k(t) e^{-(k-1)t/\sigma^2} \nonumber\\ && \qquad\qquad\qquad \int_0^t dt'\, e^{(k-1)t'/\sigma^2} \frac{\dot{g}(t')}{\sigma^2} \left( \int_0^{t'} ds\,  e^{-(k-2)(t'-s)/\sigma^2} \frac{\del}{\del\tau_{k-2}(s)}
 \right) \EEA
 
\item[(iii)] (contribution of the commutator $[\psi,\phi]$ for $t\not=t'$)
\BEQ  C_3(f,g)=\int dt\ (\frac{\dot{f}(t)}{\sigma^2}+ \ddot{f}(t))
 \sum_k k(k-1)\tau_k(t) \int_0^t dt'\,  e^{-(k-1)(t-t')/\sigma^2} \dot{g}(t')
 \frac{\del}{\del\tau_{k-2}(t')} \EEQ

\item[(iv)] ($\del$-contribution)
\BEQ C_4(f,g)=-\int dt\ (\frac{\dot{f}(t)}{\sigma^2}+ \ddot{f}(t)) \dot{g}(t)
 \sum_k k(k-1)\tau_k(t) \int_0^t ds\, e^{-(k-2)(t-s)/\sigma^2}  \frac{\del}{\del\tau_{k-2}(s)} \EEQ

\item[(v)] (zero-momentum contribution)
\BEA C_5(f,g) &=& -2N\int dt\ (\frac{\dot{f}(t)}{\sigma^2}+ \ddot{f}(t)) \tau_2(t) \int_0^t dt'\,  e^{-(t-t')/\sigma^2}
(\frac{\dot{g}(t')}{\sigma^2}+ \ddot{g}(t'))   \nonumber\\
&=& -2N\int dt\ (\frac{\dot{f}(t)}{\sigma^2}+ \ddot{f}(t)) \dot{g}(t) \tau_2(t) \EEA 
by integration by parts;

\item[(vi)] (zero-momentum contribution, continued)
\BEQ C_6(f,g)=2N\int dt\ (\frac{\dot{f}(t)}{\sigma^2}+ \ddot{f}(t))\dot{g}(t) \tau_2(t).\EEQ

\end{itemize}

Then one sees that $C_{1,1}+C_4=0$, $C_{1,2}+C_3=0$, $C_{1,3}+C_2=0$, $C_5+C_6=0$.
Consequently, $[L_{-1,quadr}(f),L_{-1,quadr}(g)]=0$.

\medskip
The contribution of $L_{-1,lin}$ to the bracket $[L_{-1}^f,L_{-1}^g]$ is easily computed,
$[L_{-1,lin}(f),L_{-1,lin}(g)]=0$ clearly while by integration by parts
\BEA && [L_{-1,lin}(f),L_{-1,quadr}(g)] - (f\leftrightarrow g) \nonumber\\
&&= \left[-\int dt\, ( \frac{\dot{f}}{\sigma^4}-\dddot{f})(t)
\int_0^t ds\,  e^{-(t-s)/\sigma^2} \frac{\del}{\del\tau_1(s)}, \frac{N}{2} \int dt'\, 
(\frac{\dot{g}(t')}{\sigma^2}+\ddot{g}(t')) \tau_1(t') \right] - (f\leftrightarrow g) \nonumber\\
&&= -\frac{N}{2} \left[ \int dt\, (\frac{\dot{f}}{\sigma^4}-\dddot{f})(t)
\int_0^t ds\,  e^{-(t-s)/\sigma^2} (\frac{\dot{g}(s)}{\sigma^2}+\ddot{g}(s)) \ -\ 
(f\leftrightarrow g) \right] \nonumber\\
&&= \frac{N}{2} \int dt \, (\dddot{f}\dot{g}-\dot{f}\dddot{g})(t)=\frac{N}{2} 
\int dt\, \frac{d}{dt}(\ddot{f}\dot{g}-\dot{f}\ddot{g}) =0.
\EEA

Thus finally: $[A(f),A(g)]=0$.



\section{Appendix}


We prove  here a certain number of explicit expressions given in section 4.
At some point we use in the proofs the following  {\em fundamental relations},
\BEQ {\cal P}_- f(z)=\frac{1}{z} \oint \frac{dw}{1-w/z}  f(w), \qquad 
{\cal P}_+ f(z)= \oint \frac{dw}{w(1-z/w)}  f(w)  \label{eq:P} \EEQ
if $f(z)=\sum_{n\in\Z} a_n z^n\in\C[[z^{-1},z]]$, where ${\cal P}_- f$, resp.
${\cal P}_+ f$,  is the projection onto ${\cal A}_-$ parallel to
${\cal A}_+$, resp. onto ${\cal A}_+$ parallel to ${\cal A}_-$, namely, ${\cal P}_- f(z)=\sum_{n\le -1} a_n z^n,\ {\cal P}_+ f(z)=\sum_{n\ge 0} a_n z^n$.
Differentiating with respect to $z$ we also get
\BEQ ({\cal P}_- f)'(z)=-\oint \frac{dw}{z^2(1-w/z)^2} f(w)=-\oint G_0^+(z^{-1},w)f(w)
\label{eq:P-cor1},\EEQ
 \BEQ  ({\cal P}_+ f)'(z)=
\oint \frac{dw}{w^2(1-z/w)^2} f(w)=\oint G_0^-(z,w^{-1})f(w). \label{eq:P-cor2} \EEQ 
Note that, by construction,
\BEQ ({\cal P}_{\pm} \hat{\psi}_t)(z)=\hat{\psi}_{\pm}(z,t), \qquad
 ({\cal P}_{\pm} \hat{\phi}_t)(z)=\hat{\phi}_{\pm}(z,t). \EEQ

\subsection{Explicit solution of equation of motion when $\beta=2$}

 We prove
here formula (\ref{eq:sol-eq-motion-beta=2}). Let $\tilde{w}\equiv w(t)\in \C[[w]]$, resp. $z(t)\in\C[[z]]$ be the solution
at time $t\in\R$ of the ODE $\dot{w}_t=-b(w(t))$, resp. $\dot{z}_t=-b(z(t))$ with initial condition $w(0)=w$, resp. $z(0)=z$.
Let $\tilde{K}_t(z^{-1},w):=\frac{1}{1-w(t)/z}$.
Then
\BEA \frac{\partial}{\partial t} \left( \frac{1}{z} \oint dw\,  \tilde{K}_t(z^{-1},w) \pi_0(w) \right) &=& \frac{1}{z^2} \oint dw\, \frac{\dot{w}(t)}{(1-w(t)/z)^2} \pi_0(w)
\nonumber\\ &=&
-\oint dw\, \frac{b(w(t))}{z^2(1-w(t)/z)^2} \pi_0(w) \nonumber\\
&=& -\oint \frac{d\tilde{w}}{z^2(1-\tilde{w}/z)^2} \frac{b(\tilde{w}) \pi_0(\tilde{w}(-t))}{\partial \tilde{w}/\partial w} \nonumber\\
&=& {\cal P}_- \left( \left( \frac{b(z)\pi_0(z(-t))}{\partial z/\partial z(-t)}
\right)'(z) \right) \EEA
We used (\ref{eq:P}) in the last step. Similarly, 
\BEA {\cal P}_- \left( \frac{\partial}{\partial z} \left( \frac{b(z)}{z} \oint \frac{dw}{1-w(t)/z} \, \pi_0(w) \right) \right) &=&  {\cal P}_- \left( b'(z) \oint dw\, \frac{\pi_0(w)}{z(1-w(t)/z)} - b(z) \oint dw\,
 \frac{\pi_0(w)}{z^2(1-w(t)/z)^2} \right) \nonumber\\
 &=& {\cal P}_- \left( b'(z) {\cal P}_-\left(  \frac{\pi_0(z(-t))}{\partial z/\partial z(-t)} \right) + 
 b(z) {\cal P}_- \left(  \left( \frac{\pi_0(z(-t))}{\partial z/\partial z(-t)} \right)'(z) \right)  \right)
  \nonumber\\
  &=& {\cal P}_- \left( \left( \frac{b(z)\pi_0(z(-t))}{\partial z/\partial z(-t)}
\right)'(z) \right) \EEA

\subsection{Solution of equation of motion when $\beta\not=2$}

Let us  now consider the equation of motion for $\beta\not=2$. To start with,
let
$\pi(z,t):=\exp(t\partial_z^2) \pi(z,0)\equiv \sum_{k\ge 0} \pi_k(t) z^{-k-1}$ be the image of $\pi(z,0)\equiv \sum_{k\ge 0}
\pi_k z^{-k-1}$
by the semi-group generated by $\partial_z^2$. One may check by inspection that
\BEQ \pi(z,t)=\frac{1}{z} \sum_{l=0}^{\infty} \frac{\pi_l z^{-l}}{l!} \sum_{m=0}^{\infty} \frac{(l+2m)!}{(2m)!} \left( \frac{t}{z^2}\right)^{2m}; \EEQ
in Fourier modes one gets
\BEQ \pi_k(t)=\sum_{m=0}^{\lfloor k/2\rfloor} \left( \begin{array}{c} k\\ 2m \end{array} \right) \pi_{k-2m} t^{2m}. \EEQ

Next we compute $K_t(z^{-1},w)$ in the {\em Hermite case} (see example in Section 1).
By definition $\pi(z,t)=\exp(t{\cal D}) \pi(z,0)=\frac{1}{z} \oint dw\, K_t(z^{-1},w) \pi(w,0)$, where $({\cal D}\pi)(z):= \frac{1}{\sigma^2} (\pi(z)+ z\pi'(z))
-(\frac{\beta}{2}-1)\pi''(z)$. 
In order to exponentiate the semi-group ${\cal D}$, we consider the formal series
\BEQ \rho(\zeta,t):=\sum_{k\ge 0} \frac{\pi_k(t)}{k!} \zeta^k \EEQ
related to $\pi(\zeta,t)$ by a Mellin transform. Through this non-local transform
$\partial_z$ becomes the multiplication by $-\zeta$, and the multiplication by $z$
becomes the derivative $\partial_{\zeta}$, hence ${\cal D}\equiv -(\frac{\beta}{2}-1)
\zeta^2-\frac{1}{\sigma^2} \zeta\partial_{\zeta}$ is a first-order operator.
Looking for a solution of the form $\rho(\zeta,t)\equiv e^{f(t)\zeta^2} \rho(g(t)\zeta,0)$, we get by identification
\BEQ \dot{g}=-\frac{1}{\sigma^2} g, \qquad \dot{f}=-\frac{\beta-2}{\sigma^2} f-(\frac{\beta}{2}-1) \EEQ
which can be solved straightforwardly, yielding 
\BEQ \rho(\zeta,t)=e^{f(t)\zeta^2} \rho(e^{-t/\sigma^2} \zeta,0)  \label{eq:rho} \EEQ
with $f(t)=-\frac{\sigma^2}{2} (1-e^{-(\beta-2)t/\sigma^2}).$
Inverting now the Mellin transform, we remark that $\rho(\zeta,t)$ is given in
(\ref{eq:rho}) as the image by $\exp(f(t)\zeta^2)\equiv \exp(f(t)\partial_z^2)$ of
a transformed initial condition $\tilde{\rho}(e^{-t/\sigma^2} \zeta)=\sum_{k\ge 0}
e^{-kt/\sigma^2}\frac{\pi_k}{k!}\zeta^k$, associated to $\tilde{\pi}(z,0)=\sum_{k\ge 0}
e^{-kt/\sigma^2} \pi_k z^{-k-1}$. Hence we get
\BEQ \pi(z,t)=\frac{1}{z} \sum_{k\ge 0} e^{-kt/\sigma^2} \frac{\pi_k z^{-k}}{k!}
\sum_{m\ge 0} \left(\begin{array}{c} k+2m\\ 2m\end{array}\right) \left( \frac{f(t)}{z^2}\right)^{2m}\EEQ
from which we finally obtain an explicit formula for $K_t$ in the Hermite case,
\BEQ \hat{K_t}(z^{-1},w)=\sum_{k\ge 0} (e^{-t/\sigma^2}w/z)^k \sum_{m\ge 0} \left(\begin{array}{c} k+2m\\ 2m\end{array}\right) \left( \frac{f(t)}{z^2}\right)^{2m}
\label{eq:Hermite-betanot=2} \EEQ
extending (\ref{eq:Hermite}). In Fourier modes this is
\BEQ \pi_k(t)=\sum_{m=0}^{\lfloor k/2\rfloor} e^{-(k-2m)t/\sigma^2} \left(
\begin{array}{c} k\\ 2m\end{array}\right) (f(t))^{2m} \pi_{k-2m}.\EEQ

\subsection{A technical lemma}

Let $u\in {\cal A}_+$. We prove here the following result,
\BEA &&  \partial_{t'} \left( \oint \frac{dw}{w}\, u(w) \partial_w(K_{t-t'}(z^{-1},w))
K_{t'-s}(w^{-1},\zeta) \right)=  \nonumber\\
&& \qquad \qquad =   \oint \frac{dw}{w} \, (u(w)b'(w)-u'(w) b(w)) \partial_w(K_{t-t'}(z^{-1},w)) K_{t'-s}(w^{-1},\zeta) \nonumber\\ \label{eq:technical-lemma} \EEA
Namely,
\BEA &&  \partial_{t'} \left( \partial_w(K_{t-t'}(z^{-1},w))\ 
K_{t'-s}(w^{-1},\zeta) \right) = \nonumber\\
&& \qquad \qquad = \partial_w(\partial_{t'}(K_{t-t'}(z^{-1},w))\  K_{t'-s}(w^{-1},\zeta)+ \partial_w(K_{t-t'}(z^{-1},w))) \  \partial_{t'}(K_{t'-s}(w^{-1},\zeta)) \nonumber\\  \EEA
is the sum of two terms. We use the second Kolmogorov formula (\ref{eq:Kolmogorov2}) for
the the first one, and the first Kolmogorov formula (\ref{eq:Kolmogorov1}) for the second one; using the fundamental relations (\ref{eq:P}, \ref{eq:P-cor1}, \ref{eq:P-cor2}) yields
\BEA &&  \oint \frac{dw}{w} u(w)\,  \partial_w(\partial_{t'}(K_{t-t'}(z^{-1},w)))\  K_{t'-s}(w^{-1},\zeta)= \nonumber\\
&& \qquad \qquad \qquad  \oint \frac{dw}{w} u(w)\,  \partial_w\left( b(w) \oint \frac{d\alpha}{\alpha^2 (1-w/\alpha)^2} K_{t-t'}(z^{-1},\alpha) \right) K_{t'-s}(w^{-1},\zeta)
\nonumber\\ && \qquad =
\oint \frac{dw}{w} u(w) \partial_w\left( b(w) \partial_w(K_{t-t'}(z^{-1},w)) \right) 
K_{t'-s}(w^{-1},\zeta) \EEA
and
\BEA && \oint \frac{dw}{w} u(w)\, \partial_w(K_{t-t'}(z^{-1},w)) \  \partial_{t'}(K_{t'-s}(w^{-1},\zeta)) = \nonumber\\
&& \qquad\qquad - \oint \frac{dw}{w} u(w)\, \partial_w(K_{t-t'}(z^{-1},w)) \ 
\oint \frac{d\alpha}{w(1-\alpha/w)^2} \frac{b(\alpha)}{\alpha} K_{t'-s}(\alpha^{-1},\zeta) \nonumber\\
&& \qquad =  \oint dw\,  u(w)\, \partial_w(K_{t-t'}(z^{-1},w))) \ 
 {\cal P}_- \left(\left( \frac{b(w)}{w} K_{t'-s}(w^{-1},\zeta) \right) \right)'(w)
\nonumber\\
&&  \qquad =  \oint dw\,  u(w)\, \partial_w(K_{t-t'}(z^{-1},w))) \ 
 \partial_w\left( \frac{b(w)}{w} K_{t'-s}(w^{-1},\zeta) \right) 
\nonumber\\
&& \qquad =-\oint dw\, \frac{b(w)}{w} K_{t'-s}(w^{-1},\zeta)\,  \partial_w\left( 
u(w)\, \partial_w(K_{t-t'}(z^{-1},w))) \right)
 \EEA
Hence the result.


\subsection{Time derivations}


One finds in the formula (\ref{eq:4.50}) for $L_0^f$ the time-derivation $f(t)\partial_t$.
By definition, it acts on local functionals of $\{(\tau_k(t))_{t\ge 0}\}_{k\in\Z}$ as an infinitesimal change of coordinates,
\BEQ f(t)\partial_t \, \cdot\, \int F(s,(\tau_k(s))_k)ds:= \sum_{k\ge 0}
\int f(s) \frac{\partial}{\partial  y_k}F(s,(y_l)_l)\big|_{\vec{y}=\vec{\tau}(s)} \, \dot{\tau}_k(s)\, ds.\EEQ
In particular, for a linear functional, one finds
\BEQ f(t)\partial_t\, \cdot\, \int g(s)\tau_k(s)\, ds =\int g(s)f(s)\dot{\tau}_k(s)\, ds
=- \int \frac{d}{ds}(f(s)g(s)) \tau_k(s)\, ds. \label{eq:app-fdtgtau}\EEQ

This action of $f(t)\partial_t$ extends in a natural way (by duality) to an action
on local functionals of $(\tau_k(\, \cdot\, ))_{k\ge 0}$ and $(\frac{\del}{\del\tau_k(\, \cdot\, )})_{k\ge 0}$. Restricting to local functionals, we impose
\BEA  0 &\equiv& f(t)\partial_t\, \cdot\,  \left( \Big\langle \int \gamma(s)\partial/\partial\tau_k(s)
\, ds, \int g(s) \tau_k(s) \,ds \Big\rangle \right) \nonumber\\
&=& \Big\langle f(t)\partial_t\, \cdot\, \int  \gamma(s)\partial/\partial\tau_k(s)
\, ds, \int g(s) \tau_k(s) \,ds \Big\rangle + 
\Big\langle \int  \gamma(s)\partial/\partial\tau_k(s)
\, ds,  f(t)\partial_t\, \cdot\, \int g(s) \tau_k(s) \,ds \Big\rangle\nonumber\\ \EEA
so
\BEQ f(t)\partial_t\, \cdot\, \int  \gamma(s)\partial/\partial\tau_k(s)
\, ds = -\int f(s) \dot{\gamma}(s) \partial/\partial\tau_k(s)\, ds. \label{eq:app-fdtgammaddtau} \EEQ

In particular,
\BEA && f(t)\partial_t \, \cdot\, \hat{\psi}_-(z,t)=-\beta^{1/2} \, \cdot\, \frac{1}{z}
\oint d\zeta \int ds\,  f(s) \frac{\partial}{\partial s} ({\bf 1}_{[0,t]}(s) K_{t-s}(z^{-1},
\zeta)) \widehat{\partial/\partial \tau}(\zeta,s) \nonumber\\
&& \ \ =\beta^{1/2} \left\{ f(t) \widehat{\partial/\partial \tau}(z,t) + \frac{1}{z}
\int_0^t ds\, f(s) \oint d\zeta \, \partial_t(K_{t-s}(z^{-1},\zeta))  \widehat{\partial/\partial \tau}(\zeta,s) \right\},  \nonumber\\  \label{eq:f(t)dtpsi-} \EEA
compare with the straightforward time-derivative formula (\ref{eq:time-derivative-psi}). Obviously the two formulas coincide when $f\equiv 1$. 

Using the semi-group property of the kernel $K$, we may express (\ref{eq:f(t)dtpsi-})
somewhat differently. First
\BEQ  f(t)\partial_t \, \cdot\, \hat{\psi}_-(z,t)= 
f(t) (\partial_t\hat{\psi}_-)(z,t)+ \beta^{1/2} \frac{1}{z} \int_0^t ds (f(s)-f(t))  \oint d\zeta \, \partial_t(K_{t-s}(z^{-1},\zeta))  \widehat{\partial/\partial \tau}(\zeta,s).\EEQ
Then, by (\ref{eq:time-derivative-psi}),
\BEA && (\partial_t\hat{\psi}_-)(z,t)=\hat{\phi}_-(z,t)+\beta^{1/2} \frac{1}{z}
\int_0^t ds \oint d\zeta \, \partial_t(K_{t-s}(z^{-1},\zeta)) \widehat{\partial/\partial
\tau}(\zeta,s) \nonumber\\
&&= \hat{\phi}_-(z,t)-\beta^{1/2}\frac{1}{z} \int_0^t ds \oint d\zeta \oint \frac{d\alpha}{\alpha^2(1-\zeta/\alpha)^2} K_{t-s}(z^{-1},\alpha) b(\zeta) \widehat{\partial/\partial \tau}(\zeta,s) \ {\mathrm{using\,}}  (\ref{eq:Kolmogorov2}) \nonumber\\
&&=  \hat{\phi}_-(z,t)+\beta^{1/2} \frac{1}{z} \int_0^t ds \oint d\alpha \, K_{t-s}(z^{-1},\alpha) \left({\cal P}_-\left(b(\zeta)\widehat{\partial/\partial\tau}(\zeta,s)\right)\right)'(\zeta=\alpha) \nonumber\\
&&= \hat{\phi}_-(z,t)-\beta^{1/2} \int_0^t ds \oint d\zeta\, G^+_{t-s}(z^{-1},\zeta)
b(\zeta) \widehat{\partial/\partial\tau}(\zeta,s)  \label{eq:app-dtpsi-bis} \EEA

and 
\BEA && \frac{1}{z} \beta^{1/2} \int_0^t ds\,  (f(s)-f(t))  \oint d\zeta \, \partial_t(K_{t-s}(z^{-1},\zeta))  \widehat{\partial/\partial \tau}(\zeta,s) \nonumber\\
&& =-\frac{1}{z} \beta^{1/2}  \int_0^t ds \, \dot{f}(s) \int_0^s dt'\, \oint d\zeta\, \partial_t (K_{t-t'}(z^{-1},\zeta)) \widehat{\partial/\partial\tau}(\zeta,t') \nonumber\\
&& =-\frac{1}{z} \beta^{1/2}  \int_0^t ds \, \dot{f}(s) \int_0^s dt'\, \oint d\zeta \oint \frac{d\xi}{\xi}
\, \partial_t(K_{t-s}(z^{-1},\xi)) K_{s-t'}(\xi^{-1},\zeta) \widehat{\partial/\partial\tau}(\zeta,t')\ {\mathrm{using\ }} (\ref{eq:KKK}) \nonumber\\
&&=  -\frac{1}{z} \int_0^t ds \, \dot{f}(s) \oint d\xi\, \partial_t( K_{t-s}(z^{-1},\xi)
) ({\cal P}_- \hat{\psi}_s)(\xi) \nonumber\\
&&= - \frac{1}{z}\int_0^t ds \, \dot{f}(s) \oint d\xi\,  \partial_t(K_{t-s}(z^{-1},\xi)) \hat{\psi}_s(\xi) \nonumber\\
&&= \frac{1}{z} \int_0^t ds\, \dot{f}(s) \oint d\xi\,  \hat{\psi}_s(\xi) \oint \frac{d\zeta}{\zeta^2(1-\xi/\zeta)^2} b(\xi) K_{t-s}(z^{-1},\zeta) \ {\mathrm{using\ }} (\ref{eq:Kolmogorov2})
\nonumber\\
&&= -\frac{1}{z} \int_0^t ds\, \dot{f}(s) \oint d\zeta \, K_{t-s}(z^{-1},\zeta) ({\cal P}_-
(b\hat{\psi}_s))'(\zeta) \nonumber\\
&&= \int_0^t ds \, \dot{f}(s) \oint d\zeta\, b(\zeta) G^+_{t-s}(z^{-1},\zeta) \hat{\psi}_s(\zeta). \label{eq:f(t)dtpsi-2} \EEA


\end{document}